\documentclass[twocolumn,tighten,times]{aastex631}

\usepackage{amsmath}
\usepackage{amssymb}
\let\tablenum\relax
\usepackage[separate-uncertainty=true,detect-weight]{siunitx}
\usepackage{cleveref}
\usepackage{microtype}
\usepackage{booktabs}
\usepackage{enumerate}
\usepackage{xcolor}

\newcommand{\hst}{\textit{HST}}
\newcommand{\jwst}{\textit{JWST}}
\newcommand{\obj}{NGC{\,}{628}}
\newcommand{\hla}{\textit{HLA}}

\newcommand{\tinytim}{\textsc{TinyTim}}
\newcommand{\imfit}{\textsc{Imfit}}
\newcommand{\astrodrizzle}{\textsc{AstroDrizzle}}
\newcommand{\webbpsf}{\textsc{WebbPSF}}

\DeclareSIUnit \pc {pc}
\DeclareSIUnit \parsec {parsec}
\DeclareSIUnit \asec {arcsec}
\DeclareSIUnit \pixel {pixel}
\DeclareSIUnit \pixels {pixels}
\DeclareSIUnit \Msun {M_{\odot}}
\DeclareSIUnit \Lsun {L_{\odot}}
\DeclareSIUnit \smass {M_{\star}}
\DeclareSIUnit \dex {dex}
\DeclareSIUnit \mag {mag}
\DeclareSIUnit \pixel {pixel}
\DeclareSIUnit \jansky {Jy}
\DeclareSIUnit \stern {sr}
\DeclareSIUnit \yr {yr}

\creflabelformat{equation}{#2#1#3}

\newcommand {\cta}[1] {\citetalias{#1}}

\defcitealias {georgiev2014a} {GB14}

\received{XXX}
\revised{YYY}
\accepted{ZZZ}

\submitjournal{ApJL}

\shorttitle{{\hst} and {\jwst} analysis of the NSC in {\obj}}
\shortauthors{Hoyer et al.}
\graphicspath{{./}}

\begin{document}

\title{PHANGS-JWST First Results: A combined {\hst} and {\jwst} analysis of the nuclear star cluster in {\obj}}

\correspondingauthor{Nils Hoyer}
\email{hoyer@mpia.de, nils.hoyer@dipc.org}

\author[0000-0001-8040-4088]{Nils~Hoyer}
\affiliation{Donostia International Physics Center, Paseo Manuel de Lardizabal 4, E-20118 Donostia-San Sebasti{\'{a}}n, Spain}
\affiliation{Max-Planck-Institut f{\"{u}}r Astronomie, K{\"{o}}nigstuhl 17, D-69117 Heidelberg, Germany}
\affiliation{Universit{\"{a}}t Heidelberg, Seminarstrasse 2, D-69117 Heidelberg, Germany}

\author[0000-0001-5965-3530]{Francesca~Pinna}
\affiliation{Max-Planck-Institut f{\"{u}}r Astronomie, K{\"{o}}nigstuhl 17, D-69117 Heidelberg, Germany}

\author[0000-0001-8768-4510]{Albrecht~W.~H.~Kamlah}
\affiliation{Astronomisches Rechen-Institut, Zentrum f{\"{u}}r Astronomie, M{\"{o}}nchhofstrasse 12-14, D-69120 Heidelberg, Germany}
\affiliation{Max-Planck-Institut f{\"{u}}r Astronomie, K{\"{o}}nigstuhl 17, D-69117 Heidelberg, Germany}

\author[0000-0002-6379-7593]{Francisco~Nogueras-Lara}
\affiliation{Max-Planck-Institut f{\"{u}}r Astronomie, K{\"{o}}nigstuhl 17, D-69117 Heidelberg, Germany}

\author[0000-0002-0160-7221]{Anja~Feldmeier-Krause}
\affiliation{Max-Planck-Institut f{\"{u}}r Astronomie, K{\"{o}}nigstuhl 17, D-69117 Heidelberg, Germany}

\author[0000-0002-6922-2598]{Nadine~Neumayer}
\affiliation{Max-Planck-Institut f{\"{u}}r Astronomie, K{\"{o}}nigstuhl 17, D-69117 Heidelberg, Germany}

\author[0000-0001-6113-6241]{Mattia~C.~Sormani}
\affiliation{Universit{\"{a}}t Heidelberg, Zentrum f{\"{u}}r Astronomie, Institut f{\"{u}}r Theoretische Astrophysik, Albert-Ueberle-Strasse 2, D-69120 Heidelberg, Germany}

\author[0000-0003-0946-6176]{Médéric~Boquien}
\affiliation{Centro de Astronomía (CITEVA), Universidad de Antofagasta, Avenida Angamos 601, Antofagasta, Chile}

\author[0000-0002-6155-7166]{Eric~Emsellem}
\affiliation{European Southern Observatory, Karl-Schwarzschild-Strasse 2, D-85748 Garching, Germany}
\affiliation{Univ Lyon, Univ Lyon1, ENS de Lyon, CNRS, Centre de Recherche Astrophysique de Lyon UMR5574, F-69230 Saint-Genis-Laval France}

\author[0000-0003-0248-5470]{Anil~C.~Seth}
\affiliation{Department of Physics and Astronomy, University of Utah, Salt Lake City, UT 84112, USA}

\author[0000-0002-0560-3172]{Ralf~S.~Klessen}
\affiliation{Universit{\"{a}}t Heidelberg, Zentrum f{\"{u}}r Astronomie, Institut f{\"{u}}r Theoretische Astrophysik, Albert-Ueberle-Strasse 2, D-69120 Heidelberg, Germany}
\affiliation{Universit{\"{a}t} Heidelberg, Interdisziplin{\"{a}}res Zentrum f{\"{u}}r Wissenschaftliches Rechnen, Im Neuenheimer Feld 205, D-69120 Heidelberg, Germany}

\author[0000-0002-0786-7307]{Thomas~G.~Williams}
\affiliation{Sub-department of Astrophysics, Department of Physics, University of Oxford, Keble Road, Oxford OX1 3RH, UK}
\affiliation{Max-Planck-Institut f{\"{u}}r Astronomie, K{\"{o}}nigstuhl 17, D-69117 Heidelberg, Germany}

\author[0000-0003-3933-7677]{Eva~Schinnerer}
\affiliation{Max-Planck-Institut f{\"{u}}r Astronomie, K{\"{o}}nigstuhl 17, D-69117 Heidelberg, Germany}

\author[0000-0003-0410-4504]{Ashley.~T.~Barnes}
\affiliation{Argelander-Institut f{\"{u}}r Astronomie, Universit{\"{a}}t Bonn, Auf dem H{\"{u}}gel 71, D-53121, Bonn, Germany}

\author[0000-0002-2545-1700]{Adam~K.~Leroy}
\affiliation{Department of Astronomy, The Ohio State University, 140 West 18th Avenue, Columbus, Ohio 43210, USA}

\author[0000-0002-6381-2052]{Silvia~Bonoli}
\affiliation{Donostia International Physics Center, Paseo Manuel de Lardizabal 4, E-20118 Donostia-San Sebasti{\'{a}}n, Spain}
\affiliation{IKERBASQUE, Basque Foundation for Science, E-48013 Bilbao, Spain}

\author[0000-0002-8804-0212]{J.~M.~Diederik~Kruijssen}
\affiliation{Cosmic Origins Of Life (COOL) Research DAO, coolresearch.io}

\author[0000-0002-3289-8914]{Justus Neumann}
\affiliation{Max-Planck-Institut f{\"{u}}r Astronomie, K{\"{o}}nigstuhl 17, D-69117 Heidelberg, Germany}

\author[0000-0003-0651-0098]{Patricia S{\'{a}}nchez-Bl{\'{a}}zquez}
\affiliation{Departamento de Física de la Tierra y Astrofísica, Universidad Complutense de Madrid, E-28040 Madrid, Spain}
\affiliation{Instituto de Física de Partículas y del Cosmos (IPARCOS), Universidad Complutense de Madrid, E-28040 Madrid, Spain}

\author[0000-0002-5782-9093]{Daniel~A.~Dale}
\affiliation{Department of Physics and Astronomy, University of Wyoming, Laramie, WY 82071, USA}

\author[0000-0002-7365-5791]{Elizabeth~J.~Watkins}
\affiliation{Astronomisches Rechen-Institut, Zentrum f{\"{u}}r Astronomie, M{\"{o}}nchhofstrasse 12-14, D-69120 Heidelberg, Germany}

\author[0000-0002-8528-7340]{David~A.~Thilker}
\affiliation{Department of Physics and Astronomy, The Johns Hopkins University, Baltimore, MD 21218, USA}

\author[0000-0002-5204-2259]{Erik~Rosolowsky}
\affiliation{Department of Physics, University of Alberta, Edmonton, Alberta, T6G 2E1, Canada}

\author[0000-0003-0166-9745]{Frank~Bigiel}
\affiliation{Argelander-Institut f{\"{u}}r Astronomie, Universit{\"{a}}t Bonn, Auf dem H{\"{u}}gel 71, D-53121 Bonn, Germany}

\author[0000-0002-3247-5321]{Kathryn~Grasha}
\affiliation{Research School of Astronomy and Astrophysics, Australian National University, Canberra, ACT 2611, Australia}   
\affiliation{ARC Centre of Excellence for All Sky Astrophysics in 3 Dimensions (ASTRO 3D), Australia}   

\author[0000-0002-4755-118X]{Oleg~V.~Egorov}
\affiliation{Astronomisches Rechen-Institut, Zentrum f{\"{u}}r Astronomie, M{\"{o}}nchhofstrasse 12-14, D-69120 Heidelberg, Germany}

\author[0000-0001-9773-7479]{Daizhong~Liu}
\affiliation{Max-Planck-Institut f{\"{u}}r Extraterrestrische Physik (MPE), Giessenbachstrasse 1, D-85748 Garching, Germany}

\author[0000-0002-4378-8534]{Karin~M.~Sandstrom}
\affiliation{Department of Physics, University of California, San Diego, 9500 Gilman Drive, San Diego, CA 92093, USA}

\author[0000-0003-3917-6460]{Kirsten L. Larson}
\affiliation{AURA for the European Space Agency (ESA), Space Telescope Science Institute, 3700 San Martin Drive, Baltimore, MD 21218, USA}

\author[0000-0003-4218-3944]{Guillermo A. Blanc}
\affiliation{The Observatories of the Carnegie Institution for Science, 813 Santa Barbara St., Pasadena, CA, USA}
\affiliation{Departamento de Astronom\'{i}a, Universidad de Chile, Camino del Observatorio 1515, Las Condes, Santiago, Chile}

\author[0000-0002-8806-6308]{Hamid Hassani}
\affiliation{Department of Physics, University of Alberta, Edmonton, Alberta, T6G 2E1, Canada}



\begin{abstract}
  We combine archival {\hst} and new {\jwst} imaging data, covering the ultraviolet to mid-infrared regime, to morphologically analyze the nuclear star cluster (NSC) of {\obj}, a grand-design spiral galaxy.
  The cluster is located in a $\SI{200}{\pc} \times \SI{400}{\pc}$ cavity, lacking both dust and gas.
  We find roughly constant values for the effective radius ($r_{\mathrm{eff}} \sim \SI{5}{\pc}$) and ellipticity ($\epsilon \sim \num{0.05}$), while the S{\'{e}}rsic index ($n$) and position angle (\textit{PA}) drop from $n \sim \num{3}$ to $\sim \num{2}$ and $\mathrm{\textit{PA}} \sim \SI{130}{\degree}$ to \SI{90}{\degree}, respectively.
  In the mid-infrared, $r_{\mathrm{eff}} \sim \SI{12}{\pc}$, $\epsilon \sim \num{0.4}$, and $n \sim \num{1}$-\num{1.5}, with the same $\mathrm{\textit{PA}} \sim \SI{90}{\degree}$.
  The NSC has a stellar mass of $\log_{10} \, (M_{\star}^{\mathrm{nsc}} \, / \, \si{\Msun}) = \num{7.06} \pm \num{0.31}$, as derived through $\mathrm{\textit{B}} - \mathrm{\textit{V}}$, confirmed when using multi-wavelength data, and in agreement with the literature value.
  Fitting the spectral energy distribution, excluding the mid-infrared data, yields a main stellar population's age of $(\num{8} \pm \num{3}) \, \si{\giga\yr}$ with a metallicity of $Z = \num{0.012} \pm \num{0.006}$.
  There is no indication of any significant star formation over the last few \si{\giga\yr}.
  Whether gas and dust were dynamically kept out or evacuated from the central cavity remains unclear.
  The best-fit suggests an excess of flux in the mid-infrared bands, with further indications that the center of the mid-infrared structure is displaced with respect to the optical centre of the NSC.
  We discuss five potential scenarios, none of them fully explaining both the observed photometry and structure.
\end{abstract}



\section{Introduction}
\label{sec:introduction}

Nuclear star clusters (NSCs) are massive and compact stellar systems in galactic nuclei.
The effective radii range from a few to tens of parsecs.
Such radii are typical of globular clusters and ultra-compact dwarfs \citep[e.g.][]{georgiev2014a,norris2014a,pechetti2020a}.
Stellar masses may reach up to \SI{e9}{\Msun} \citep[e.g.][]{georgiev2016a}, which, in combination with the small effective radii, lead to core-densities that can approach $\lesssim \SI{e8}{\Msun\per\pc\cubed}$ \citep[e.g.][]{stone2017b} effectively making NSCs the densest stellar systems known \citep[see][for a review]{neumayer2020a}.

The formation and growth of NSCs depends on host galaxy mass \citep{fahrion2021a}, and potentially morphological type \citep{pinna2021a}.
Two main scenarios have been proposed in the literature: 
dissipationless globular cluster (GC) migration dominates in the dwarf galaxy regime ($M_{\star}^{\mathrm{gal}} < \SI{e9}{\Msun}$; \citealp{tremaine1975a,capuzzo-dolcetta1993a,agarwal2011a,hartmann2011a,arca-sedda2014b,antonini2015b,fahrion2020a,fahrion2022a,fahrion2022b}) and \textit{in-situ} star formation in more massive galaxies \citep{milosavljevic2004a,bekki2006a,bekki2007a,turner2012a,sanchez-janssen2019a,neumayer2020a}.
The latter scenario requires gas inflow, which may be caused by non-axisymmetric potentials (for example bars, \citealp{shlosman1990a}), dynamical friction of star-forming clumps \citep[e.g.][]{bekki2006a,bekki2007a}, supernova driven turbulence \citep[e.g.][]{sormani2020b,tress2020b}, or rotational instabilities of the disk \citep{milosavljevic2004a}.
Once the gas settles at the center of the cluster and cools off, star formation begins, leading to the observation of young stellar populations \citep[e.g.][]{walcher2006a,rossa2006a,seth2008b,kacharov2018a,hannah2021a,fahrion2021a} and structural variations, such as a wavelength-dependent effective radius \citep{georgiev2014a,carson2015a}.
Young stellar populations were also directly observed in various nuclei, including the Milky Way's NSC \citep[e.g.][]{seth2006a,seth2008b,do2009b,genzel2010b,carson2015a,feldmeier-krause2015a,kacharov2018a,nguyen2019a,hannah2021a,henshaw2022b}.
A combination of both GC migration and \textit{in-situ} star formation is also possible if the infalling GC keeps a gas reservoir and continues star formation during inspiral \citep{guillard2016a}.
Corroborated by scaling relations between cluster properties and their host galaxies \citep[e.g.][]{ferrarese2006a,seth2008b,erwin2012a,scott2013a,ordenes-briceno2018b,sanchez-janssen2019b}, studying nuclear clusters in detail reveals both their formation history as well as that of their host galaxy.
Drawing this connection in {\obj} is one of the goals of this work.

NSCs appear frequently, albeit not ubiquitously, at galaxy masses of \num{e9}{-}\SI{e10}{\Msun} in various environments \citep{cote2006a,turner2012a,baldassare2014a,denbrok2014a,neumayer2020a,hoyer2021a}.
While this fraction decreases towards higher galaxy masses, there are indications that it drops at a slower rate for late-type galaxies compared to early-types \citep{neumayer2020a,hoyer2021a}.
A majority of NSCs in high-mass galaxies were discovered in spiral galaxies \citep[e.g.][]{carollo1998a,boeker2002a}, likely due to high central luminosities of massive elliptical galaxies.

One example for a nucleated, massive ($M_{\star}^{\mathrm{gal}} \sim \SI{2e10}{\Msun}$; \citealp{leroy2021a}) galaxy is {\obj} (M{\,}74), the object of this study.
The NSC was analyzed previously by \citet[][hereafter \cta{georgiev2014a}]{georgiev2014a} using \textit{Hubble Space Telescope} ({\hst}) WFPC2 imaging data, but no in-depth analysis of all available high-resolution data has been performed yet.
With the advent of the \textit{James Webb Space Telescope} ({\jwst}) earlier this year, we aim to study the NSC of {\obj} across the optical and infrared regimes, analyzing both its structural and photometric properties.

This grand-design spiral galaxy is located at a distance of $d = \num{9.84} \pm 0.63 \, \si{\mega\pc}$ \citep[][and also \citealp{mcquinn2017a}]{anand2021a} at the edge of the Local Volume ($\lesssim \SI{11}{\mega\pc}$).
Both its relatively isolated position \citep[.g.][]{briggs1980a} and nearly face-on orientation ($i \sim \SI{8.9}{\degree}$; \citealp{lang2020a}) make the galaxy an optimal test-case for detailed studies of galactic disks, and star- and cluster-formation in massive late-types \citep[see e.g.][]{elmegreen1984a,condon1987a,grasha2015a,adamo2017a,mulcahy2017a,kreckel2018a,sun2018a,vilchez2018a,schinnerer2019a,zaragoza-cardiel2019a,chevance2020a,yadav2021a}.

\Cref{fig:overview} shows an overview of the innermost $\SI{20}{\arcsec} \times \SI{20}{\arcsec}$ (approximately $\SI{950}{\pc} \times \SI{950}{\pc}$) of {\obj}.
Corroborated by \textit{AstroSat} UV, \textit{MUSE} H$\alpha$ \citep{emsellem2022a}, and \textit{ALMA} CO maps \citep{leroy2021a}, the {\hst} and {\jwst} data reveal a spheroidal component, dust and gas reservoirs along prominent spiral arm structures, and star-forming regions.
Instead of continued spiral arms down to the smallest scales, a central cavity of approximately $\SI{200}{\pc} \times \SI{400}{\pc}$ lacking both gas and dust is present.
The NSC of {\obj} appears as a prominent bright source in the center of the galaxy.

Secular evolution plays a key role in the history of {\obj}, as indicated by the presence of a circum-nuclear region of star formation with radius $\sim \SI{25}{\arcsec}$ ($\sim \SI{1.2}{\kilo\pc}$) \citep{sanchez2011a}.
While the formation of such a region can be related to the presence of a bar \citep{piner1995a,sakamoto1999a,sheth2005a,fathi2007a,sormani2015a,spinoso2017a,bittner2020a}, as argued to be present in {\obj} by \citet{seigar2002a} and \citet{sanchez-blazquez2014a}, more recent work finds that {\obj} does not contain an obvious bar \citep{querejeta2021a}, despite an observed metallicity gradient, which is related to mixing of gas induced by a bar-structure in other late-type galaxies \citep{martin1995a,friedli1995a,dutil1999a,scarano2013a}.
If not by a bar, the presence of a circum-nuclear region of star formation may also be caused by past minor mergers, as speculated for other unbarred late-types by \citet{silchenko2006a}.
Indeed, dwarf galaxies are known to exist around {\obj} \citep{davis2021a}.
It is also plausible that the galaxy hosted a bar in the past, which was destroyed by minor mergers \citep{cavanagh2022a}.

The large, approximately $\SI{200}{\pc} \times \SI{400}{\pc}$ large, central cavity remains challenging to explain.
Currently, it is unclear whether inflow of gas and dust is prohibited dynamically, or whether the material has been expelled by star formation, supernovae, or a previously accreting massive black hole.
As motivated above, the NSC properties may give inform us on the evolution of {\obj}, if studied in detail.
Therefore, one of the goals of this study is to relate the NSC properties to the evolution of its host galaxy.

In this work, we combine archival {\hst} and newly obtained {\jwst} imaging data to study the NSC of {\obj} in great detail.
Our data extend from the ultraviolet to the mid-infrared regime (see \Cref{fig:transmission} and \Cref{tab:data_overview}) and are of high-enough resolution to resolve the cluster at all wavelengths.
This study presents the first analysis of an NSC with {\jwst} data and highlights the telescope's scientific value for studies of galactic nuclei in the local Universe.
Using the available data, we derive photometric and structural parameters for all bands, and model the spectral energy distribution of the NSC.

We introduce the data from both space telescopes in Section~\ref{sec:data} and briefly discuss the data processing pipelines as well as the generation of synthetic point spread functions.
Image analysis is described in Section~\ref{sec:image_fitting} and the main analysis steps are detailed in Section~\ref{sec:analysis}.
The results of the study are discussed in Section~\ref{sec:discussion}.
We conclude in Section~\ref{sec:conclusions}.
\begin{figure*}
  \centering
  \includegraphics[width=\textwidth]{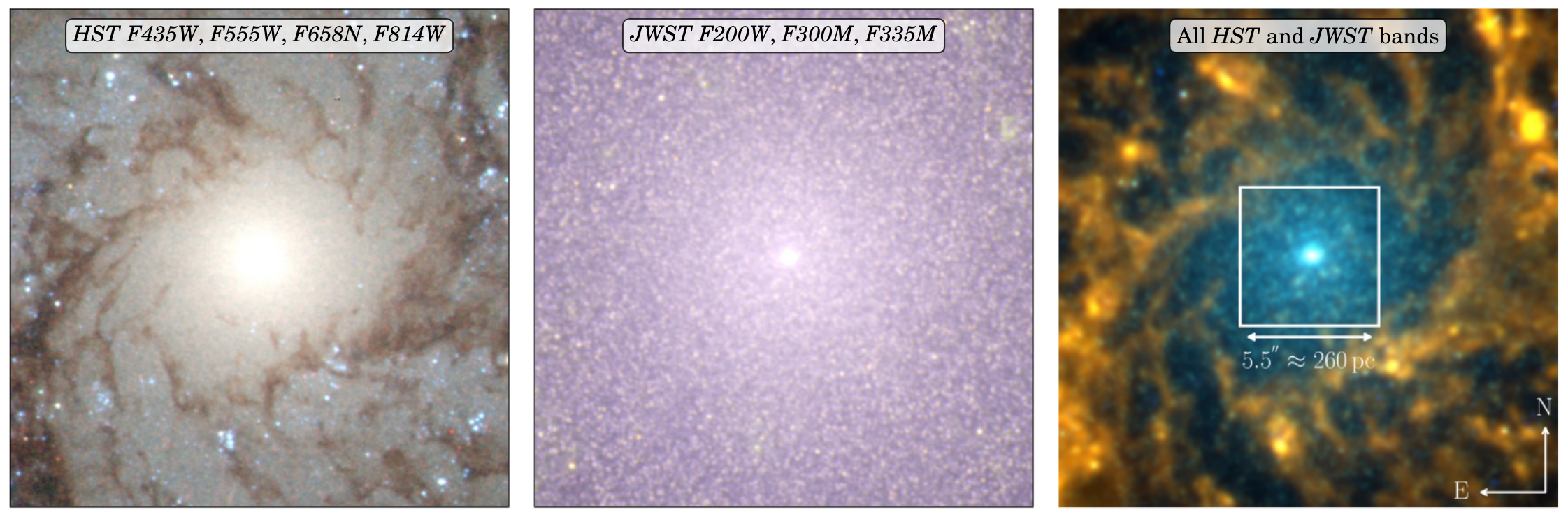}
  \includegraphics[width=\textwidth]{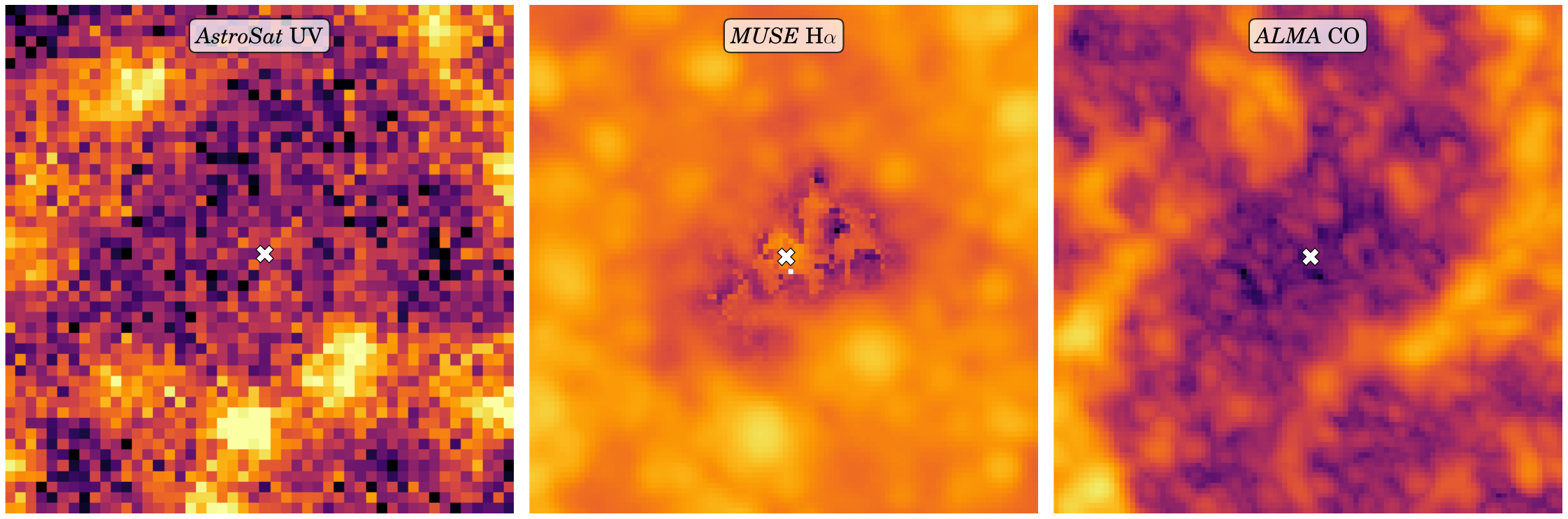}
  \includegraphics[width=\textwidth]{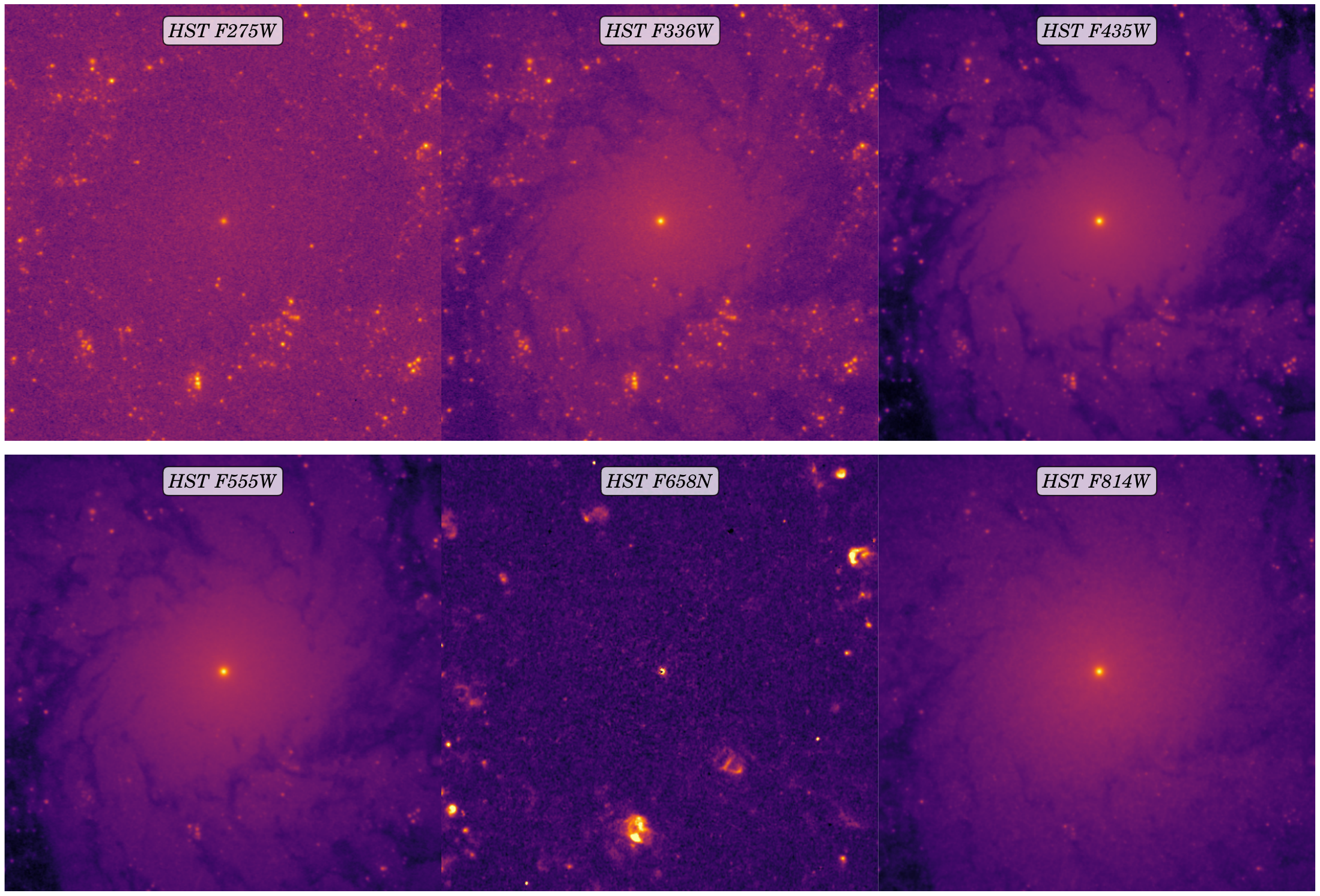}
\end{figure*}

\begin{figure*}
  \centering
  \includegraphics[width=\textwidth]{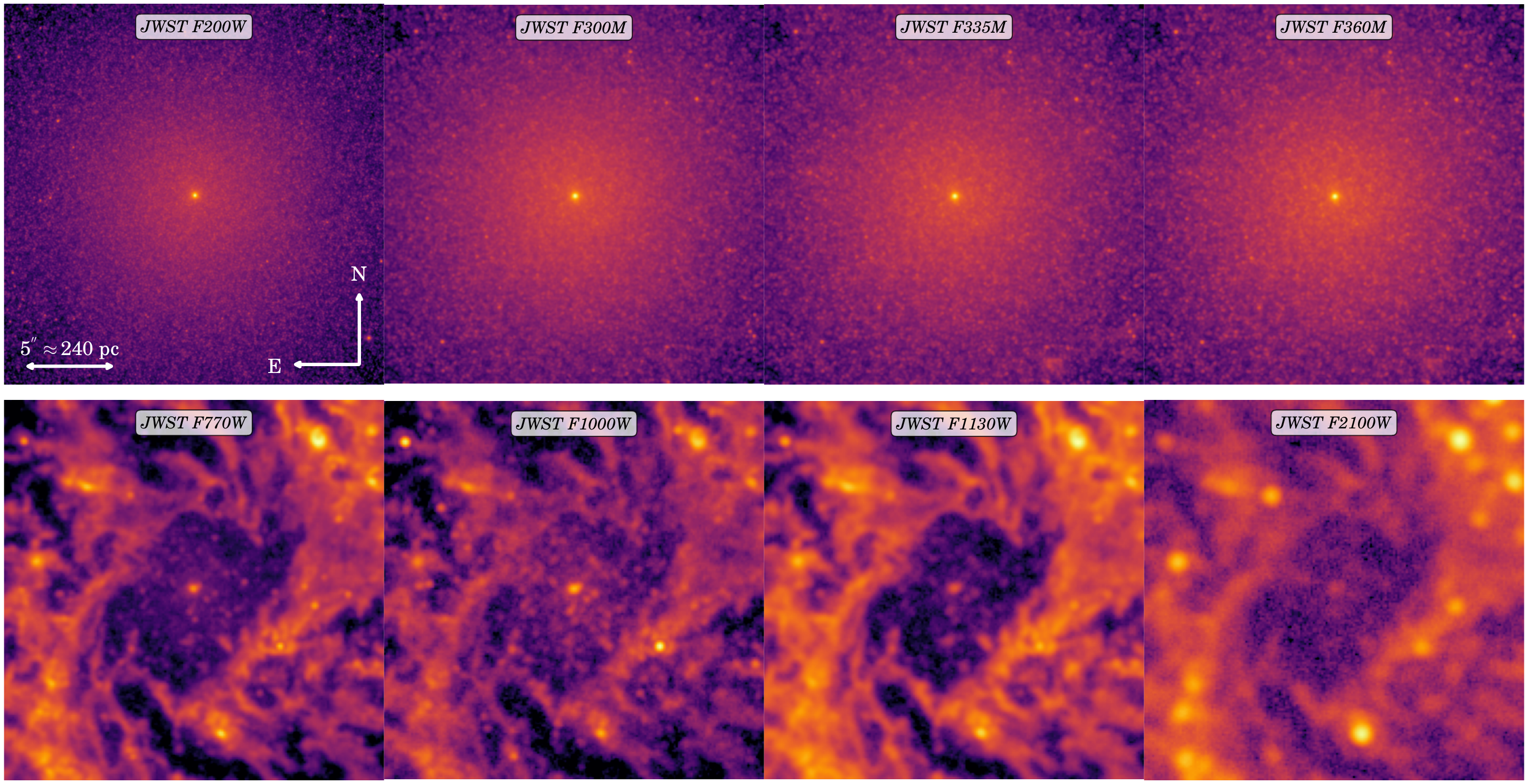}
  \caption{%
    Overview of the nuclear region of {\obj} in different bands.
    Each panel gives the central $\SI{20}{\arcsec} \times \SI{20}{\arcsec}$ (approximately $\SI{950}{\pc} \times \SI{950}{\pc}$) of the galaxy, centered on the nuclear star cluster.
    North is up and East is to the left.
    Color correlates with intensity.
    The \textit{first row} gives three color-images using {\hst} and {\jwst} bands, highlighting star-formation by using the continuum-subtracted {\hst} ACS \textit{F658N} (H$\alpha$) filter.
    Dust lanes, where star formation occurs, are clearly visible in the first and third panel, while only stellar emission is shown in the second panel. 
    The squared box of side length \SI{5.5}{\arcsec} in the third panel shows the region we considered for the fit of the NSC.
    The \textit{second row} highlights other available data sets, namely the \textit{AstroSat} ultraviolet emission, \textit{MUSE} H$\alpha$, and \textit{ALMA} CO.
    The first two panels trace again star formation whereas the third panel shows the location of the molecular gas.
    Note the absence of star formation and gas in the immediate vicinity of the nuclear star cluster, which is marked with a white cross.
    The \textit{third} and \textit{fourth rows} show the {\hst} data, increasing in wavelength.
    Star forming regions, identified in the {\hst} WFC3 \textit{F275W}, become hidden behind dust filaments in other bands.
    The sixth panel, showing the {\hst} ACS \textit{F814W} data, mainly shows stellar emission.
    We show the newly obtained {\jwst} NIRCam (\textit{fifth}) and MIRI (\textit{sixth}) in the last two rows, again in increasing wavelength from the near-infrared to the mid-infrared.
    The NIRCam data highlights the stellar emission while the MIRI data shows both stellar and dust emission.
    As for the molecular gas (\textit{ALMA} CO, second row), a central cavity exists and dust is present in the spiral arms of {\obj}.
  }
  \label{fig:overview}
\end{figure*}
\begin{figure*}
  \centering
  \includegraphics[width=\textwidth]{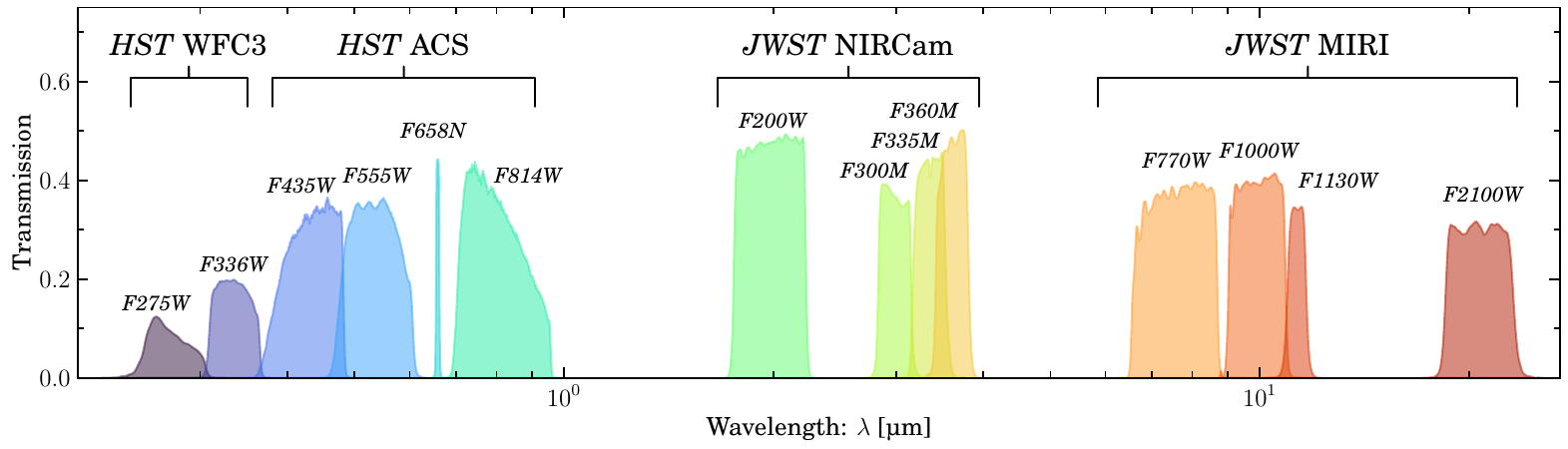}
  \caption{%
    Transmission of the {\hst} and {\jwst} bands used in this work.
  }
\label{fig:transmission}
\end{figure*}
\section{Data}
\label{sec:data}

Our analysis is based on archival {\hst} ACS \& WFC3 taken from the \textit{Hubble Legacy Archive}\footnote{\url{https://hla.stsci.edu/}} and recently obtained {\jwst} NIRCam \& MIRI imaging data (project ID 02107, PI J. Lee; see \citealp{LEE_PHANGSJWST}, this issue).
A brief overview of the available data is given in \Cref{tab:data_overview} and \Cref{fig:transmission}.
In the next three subsections we briefly describe the data processing for each instrument, followed by the generation of point spread functions.

\begin{deluxetable*}{
    l
    l
    l
    l
    l
    l
  }
  \tablecaption{%
    Data used in this work.
    \label{tab:data_overview}
  }
  \tablehead{
    \multicolumn{1}{c}{Instrument} & \multicolumn{1}{c}{Channel} & \multicolumn{1}{c}{Filter} & \multicolumn{1}{c}{PropID} & \multicolumn{1}{c}{$t_{\mathrm{exp}}$} & \multicolumn{1}{c}{pixel scale\tablenotemark{{\small a}}}\\
     & & & & \multicolumn{1}{c}{[\si{\second}]} & \multicolumn{1}{c}{[\si{\asec\per\pixel}]}
  }
  \startdata
    {\hst}  & {WFC3}   & {\textit{F275W}}  & 13364 & \num{4962.00} & \num{0.040} \\
    {\hst}  & {WFC3}   & {\textit{F336W}}  & 13364 & \num{4962.00} & \num{0.040} \\ \addlinespace \hline \addlinespace
    {\hst}  & {ACS}    & {\textit{F435W}}  & 10402 & \num{2716.00} & \num{0.050} \\
    {\hst}  & {ACS}    & {\textit{F555W}}  & 10402 & \num{1716.00} & \num{0.050} \\
    {\hst}  & {ACS}    & {\textit{F658N}}  & 10402 & \num{2844.00} & \num{0.050} \\
    {\hst}  & {ACS}    & {\textit{F814W}}  & 10402 & \num{1844.00} & \num{0.050} \\ \addlinespace \hline \addlinespace
    {\jwst} & {NIRCam} & {\textit{F200W}}  & 02107 & \num{9620.16} & \num{0.031} \\
    {\jwst} & {NIRCam} & {\textit{F300M}}  & 02107 & \num{773.048} & \num{0.063} \\
    {\jwst} & {NIRCam} & {\textit{F335M}}  & 02107 & \num{773.048} & \num{0.063} \\
    {\jwst} & {NIRCam} & {\textit{F360M}}  & 02107 & \num{858.944} & \num{0.063} \\ \addlinespace \hline \addlinespace
    {\jwst} & {MIRI}   & {\textit{F770W}}  & 02107 & \num{266.40}  & \num{0.110} \\
    {\jwst} & {MIRI}   & {\textit{F1000W}} & 02107 & \num{366.30}  & \num{0.110} \\
    {\jwst} & {MIRI}   & {\textit{F1130W}} & 02107 & \num{932.412} & \num{0.110} \\
    {\jwst} & {MIRI}   & {\textit{F2100W}} & 02107 & \num{965.712} & \num{0.110} \\
  \enddata
  \tablenotetext{{\small \textit{a}}}{Original pixel values, which remained unchanged during data processing.} 
\end{deluxetable*}

\subsection{Hubble Space Telescope}
\label{subsec:hubble_space_telescope}

We obtain all available flat-fielded single exposures from the {\hla} to combine them into a single master frame.
As a first step, the world coordinate systems of the ACS \& WFC3 received updates using the latest reference files.
These updated files were fed to {\astrodrizzle} \citep[][see also \citealp{fruchter1997a} for the drizzle algorithm]{fruchter2010a,gonzaga2012a}, which combines them into a master science product given user-specified settings.
As tested and justified in other work \citep{hoyer2022a}, we chose a pixel fraction of \num{0.75} but keep the pixel scale at their original resolutions (see \Cref{tab:data_overview}).%
\footnote{The pixel fraction controls how individual exposures are added: a value of zero corresponds to pure interlacing whereas a value of one results in a ``shift-and-add'' style of pixel values.}
No additional sky subtraction was performed as we account for background flux from the galaxy with a S{\'{e}}rsic profile \citep{sersic1968a} and a plane offset.

\subsection{James Webb Space Telescope}
\label{subsec:james_webb_space_telescope}

As part of the ``Physics at High Angular resolution in Nearby GalaxieS'' (PHANGS) {\jwst} cycle 1 treasury program, {\obj} was observed in various NIRCam and MIRI bands on July 17, 2022 (see \Cref{tab:data_overview}, and also \citealp{LEE_PHANGSJWST}, this issue).
Data reduction and co-addition were carried out using a custom data reduction pipeline, which among other things, improves the astrometric solutions and zero point offsets compared to the publicly available data products.
More specifically, the world coordinate system (WCS) was updated to match the one from the {\hst} and \textit{Gaia}, and overall background level were calibrated against e.g.\ \textit{IRAC4} \SI{8}{\micro\metre} and \textit{WISE3} \SI{12}{\micro\metre} fluxes (\citealp{LEROY1_PHANGSJWST}, this issue).
More detail of the customized version of the data reduction pipeline will be presented in \citealp{LEE_PHANGSJWST}~(this issue).

\subsection{Point spread functions}
\label{subsec:point_spread_functions}

For all bands, we used artificially generated point spread functions (PSFs) instead of determining them from non-saturated stars in the images.
The main reason for this choice was that no star is unaffected by dust and falls close to the location of the NSC (see \Cref{fig:overview}).
Especially the latter condition is important for {\hst} data as the PSF is known to vary significantly across the whole chip.

Following the approach by \citet{hoyer2022a}, PSFs were generated using {\tinytim} \citep{krist1993a,krist1995a} for {\hst} bands.
To minimize systematic differences in data processing, we did not directly use the resulting PSF from {\tinytim} for deriving the structural properties.
Instead, we copied all input science frames and set their first header extension (science data) to zero.
We then generated PSFs at the location of the NSC on each individual exposure and placed them into the previously normalized frames, taking into account the orientation of the original science images.
Afterwards, we repeated the {\astrodrizzle} processing for the normalized frames in the same way as for the science data.
The final PSF was extracted from the output of {\astrodrizzle}.
In comparison to a PSF from {\tinytim}, the core of the extracted PSF is slightly more extended due to the drizzling process.
Taking this effect into account is important for deriving accurate effective radii, as detailed in \citet{hoyer2022a}.

Generation of PSFs for {\jwst} bands was performed with {\webbpsf} \citep{perrin2012a,perrin2014a}.
To generate a star at the location of the NSC, we first generated a grid of 36 PSFs for the detector elements where the position of the NSC falls upon.
The PSF for the position of the NSC was evaluated based on interpolation of generated PSFs using {\webbpsf}.
This step is crucial as the PSF of {\jwst} varies in both the spatial and temporal dimension \citep{nardiello2022a}.
By default, and in agreement with our choice for the {\tinytim}-based PSFs, we chose a \texttt{G2V} star as the stellar template.
As explored in \citet{hoyer2022a}, the choice of stellar type plays little to no role on the fit results for {\hst} data and we assume the same for {\jwst} data.
\section{Image fitting}
\label{sec:image_fitting}

\subsection{Approach and model function}
\label{subsec:approach_and_model_function}

Focusing on the center of {\obj}, we extracted a square region of side length \SI{5.5}{\arcsecond} (equivalent to $\sim \SI{260}{\pc}$) centered on the NSC to avoid the more dust- and gas-rich spiral structure, as shown in the first row, right panel of \Cref{fig:overview}.
Previous investigations used various model functions to describe the light distribution of NSCs, including King profiles \citep[see \citealp{king1962a} for the original definition; e.g.][]{matthews1999a,seth2006a,georgiev2009a,georgiev2014a}, Gaussian profiles \citep[e.g.][]{carollo1997a,carollo2002a,barth2009a,denbrok2014a}, Nuker profiles \citep[see \citealp{lauer2005a} for a definition, e.g.][]{carollo1998a,boeker1999a,boeker2002a,butler2005a}, S{\'{e}}rsic profiles \citep[e.g.][]{cote2006a,baldassare2014a,carson2015a,spengler2017a,pechetti2020a}, or point sources \citep[e.g.][]{ferrarese2020a,poulain2021a,zanatta2021a,carlsten2022a}.
Here we use the S{\'{e}}rsic profile of the form
\begin{equation}
  \label{equ:sersic}
  I(r) = I_{\mathrm{eff}} \, \exp \bigg\{ -b_{n} \, \bigg[ \bigg( \frac{r}{r_{\mathrm{eff}}} \bigg)^{1/n} - 1 \bigg] \bigg\} \; ,
\end{equation}
where $r$ is the radius, $r_{\mathrm{eff}}$ the half-light or effective radius, $I_{\mathrm{eff}}$ the intensity at $r_{\mathrm{eff}}$, and $n$ the S{\'{e}}rsic index.
The parameter $b_{n}$ solves the equation $\Gamma (2n) = 2 \gamma(2n, \, b_{n})$ where $\gamma(a,x)$ is the incomplete and $\Gamma(x)$ the complete Gamma function.
For $n \in (\num{0.5}, \, \num{10})$, $b_{n} = \num{1.9992} n - \num{0.3271}$ is a good approximation \citep[][but see also \citealp{graham2005a} for a general overview]{capaccioli1989a}.

The background flux from the host galaxy and the sky was modeled with another S{\'{e}}rsic profile and a plane offset.
From our fits (see below), we found that in all bands $I_{\mathrm{eff}}^{\mathrm{gal}} \ll I_{\mathrm{eff}}^{\mathrm{nsc}}$, $r_{\mathrm{eff}}^{\mathrm{gal}} \gg r_{\mathrm{eff}}^{\mathrm{nsc}}$, and $n^{\mathrm{gal}} \lesssim 1$, such that the profile of the host galaxy becomes flat in the very center, thus justifying the choice of models.
As we describe in Appendix \ref{sec:number_of_sersic_profiles}, two S{\'{e}}rsic profiles describe the NSC worse than a single profile.

To fit the data, we convolved the profiles with the previously generated synthetic PSF at the position of the NSC (see Section \ref{subsec:point_spread_functions}).
The fit itself was performed with {\imfit} \citep{erwin2015a}, a specialized program to fit astronomical images.
For the minimization technique, we chose the Differential Evolution solver with Latin hypercube sampling.
In comparison with other available options, the solver does not rely on initial parameter values but randomly selects parameter values between user-specified boundaries \citep[see][for details]{storn1997a}.
We list the boundaries for the parameters of the S{\'{e}}rsic profile used for the NSC in \Cref{tab:sersic_parameters}. 
By default, {\imfit} evaluates the goodness of the fit with standard $\chi^{2}$ statistics.

Unless specified, {\imfit} assumes Poissonian statistics of the input data to generate a noise map.
We take this approach for all but the {\jwst} MIRI data where noise maps were generated by the previously mentioned custom data calibration pipeline.
The noise maps for the MIRI bands were determined from uncertainties of the input data, the read noise and the flat fields, weighted by the fractional contribution to each pixel.
As a result, compared to the standard noise map generated by {\imfit}, the MIRI noise maps give lower values for the nucleus itself, but higher values for the faint emission by the background galaxy.

\begin{figure*}
  \centering
  \includegraphics[width=\textwidth]{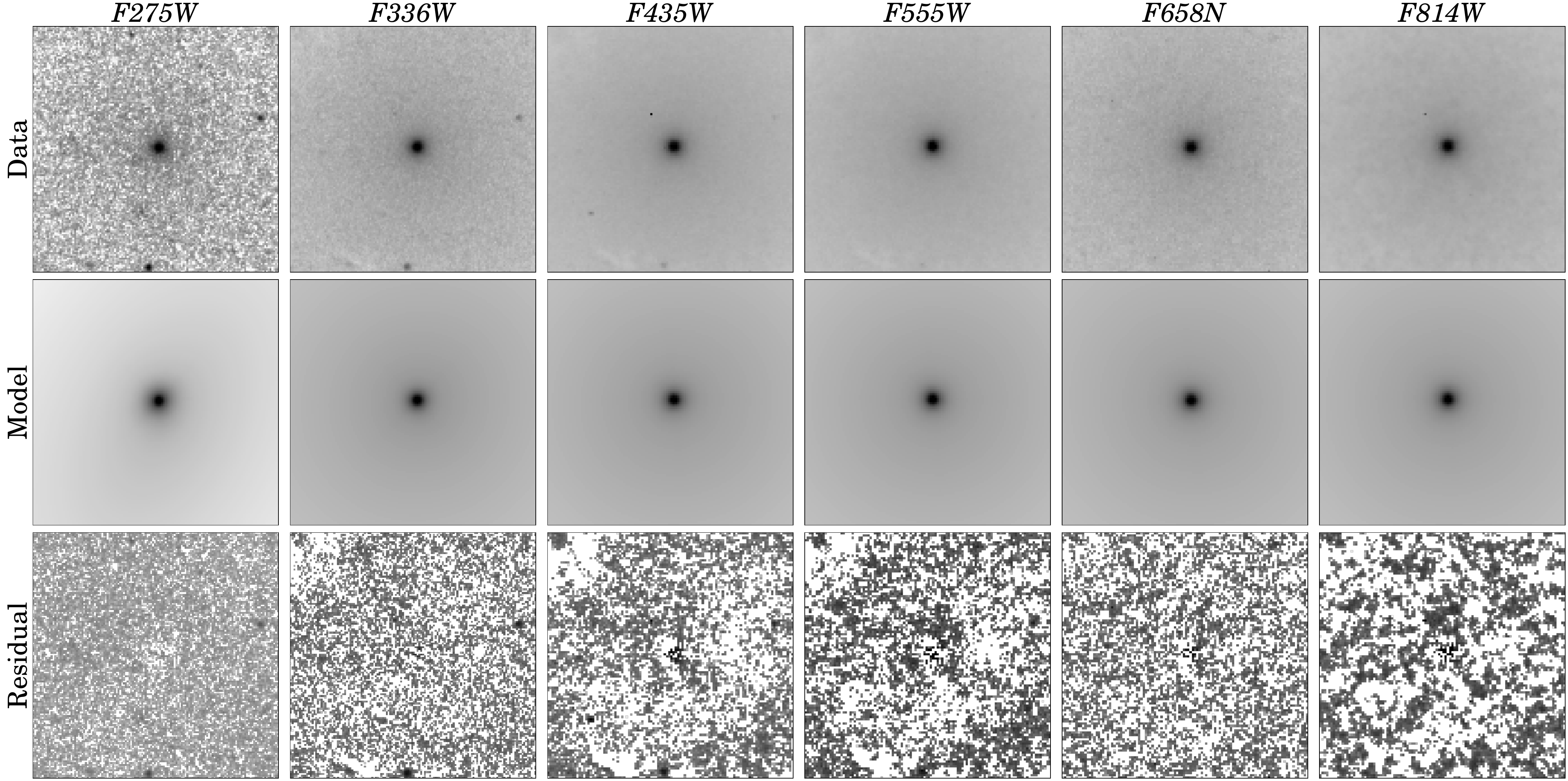}
  \caption{%
    Overview of the central $\SI{5.5}{\arcsec} \times \SI{5.5}{\arcsec}$ ($\SI{1}{\arcsec} \approx \SI{47}{\pc}$) of {\obj} in six different \textit{Hubble Space Telescope} bands.
    North is up and East is left.
    \textit{Top row}: Data products as used for fitting.
    \textit{Middle row}: Model images of the best fit using one
    S{\'{e}}rsic profile for the nuclear star cluster, an additional S{\'{e}}rsic profile for the background, and a plane offset.
    \textit{Bottom row}: Residual map, $\mathrm{Data} - \mathrm{Model}$.
    The gray scale of the maps are logarithmic and vary from \num{5e-3} to \num{5e2} times the mean of the residuals.
  }
  \label{fig:fit_hst}
\end{figure*}

\begin{figure*}
  \centering
  \includegraphics[width=\textwidth]{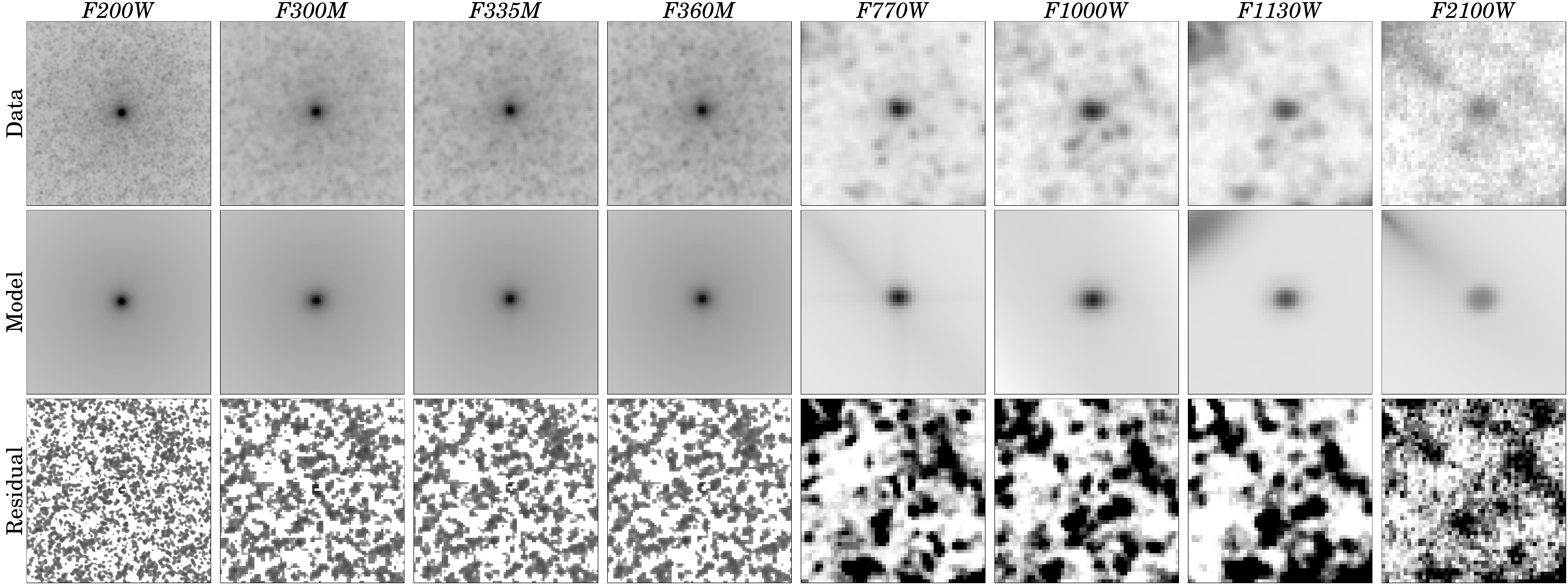}
  \caption{%
    Overview of the same central region as in \Cref{fig:fit_hst} but for available {\jwst} NIRCam and MIRI data.
    For the \textit{F2100W}, no fit was possible including PSF convolution.
    The model shown here is a pure S{\'{e}}rsic profile without PSF convolution to determine the apparent magnitude of the source.
    For the MIRI data, we adjust the gray scales to \num{0.99} and \num{1.01} times the mean.
  }
  \label{fig:fit_jwst}
\end{figure*}

In \Cref{fig:fit_hst,fig:fit_jwst} we show the data, best-fit two-component models, and the residuals for the {\hst} ACS \& WFC3, {\jwst} NIRCam, and MIRI data, respectively.
For the {\jwst} MIRI \textit{F2100W} all attempts failed to find a stable fit.
Instead, to get an estimate for the apparent magnitude, we fit a S{\'{e}}rsic profile excluding PSF convolution.
\begin{deluxetable}{
    l
    l
    l
    l
  }
  \tablecaption{%
    S{\'{e}}rsic parameters and their boundary values used to fit the data with {\imfit}.
    The same values are used for all {\hst} and {\jwst} bands.
    \label{tab:sersic_parameters}
  }
  \tablehead{%
    \multicolumn{1}{c}{Parameter} & \multicolumn{1}{c}{Boundary} & \multicolumn{1}{c}{Unit} & \multicolumn{1}{c}{Description}
  }
  \startdata
    $x_{0}$            & $[\num{45}, \, \num{55}]$                             & \si{\pixel}  & NSC position \\
    $y_{0}$            & $[\num{45}, \, \num{55}]$                             & \si{\pixel}  & NSC position \\
    PA                 & $[\num{-359.99}, \, \num{359.99}]$\tablenotemark{{\small a}}         & \si{\deg}    & Position angle \\
    $\epsilon$         & $[\num{0.00}, \, \num{0.99}]$                         & {-{-}}       & Ellipticity \\
    $n$                & $[\num{0.00}, \, \num{5.00}]$                         & {-{-}}       & S{\'{e}}rsic index \\
    $r_{\mathrm{eff}}$ & $[\num{0.00}, \, \num{10.00}]$                        & \si{\pixel}  & Effective radius \\
    $I_{\mathrm{eff}}$ & $[\num{0.00}, \, I_{\mathrm{max}}]$\tablenotemark{{\small b}}  & counts       & Intensity at $r_{\mathrm{eff}}$
  \enddata
  \tablenotetext{a}{%
    To prevent the fit from running into boundaries at \SI{0}{\degree}, the lower boundary was changed to negative values.
    In the case that the best-fit position angle was negative, \SI{180}{\degree} (or \SI{360}{\degree}) was added.
  }
  \tablenotetext{b}{%
    $I_{\mathrm{max}}$ is the peak of the intensity of the NSC.
  }
\end{deluxetable}


\subsection{Uncertainties}
\label{subsec:uncertainties}

We determined uncertainties by repeating the fit \num{500} times using bootstrapping.
During each iteration of bootstrapping, {\imfit} generates a new data array where indices of pixels are randomly sampled.
During re-sampling, the pixel values of the input data as well as their location are not considered.
The quoted best-fit parameters equal the median value of the parameter distribution and the uncertainties give the $\num{1} \sigma$ interval.

For some physical parameters, such as the determination of NSC mass (see Section \ref{subsec:stellar_mass}) or the transformation of the effective radius to parsec, the bootstrapping uncertainties were propagated forward.
Based on the assumption that the underlying probability distributions are Gaussians, we used the Gaussian error propagation.

The uncertainty of the zero point magnitudes for the {\hst} bands is of the order $\mathcal{O} (\num{10}^{\num{-3}})$ and can be neglected.
However, recent analyses of early {\jwst} data revealed that there exist issues with the flux calibration.
As detailed below, these issues persist and remain significant

For the {\jwst} NIRCam data, the uncertainty on the flux calibration can be as high as \SI{0.2}{\mag} \citep{boyer2022a}, depending on the band and detector.
Most recent analyses in the community try to solve this issue by introducing multiplicative correction factors for the data.\footnote{See, for example, \url{https://github.com/gbrammer/grizli/pull/107}.}
We corrected the determined fluxes by the mean multiplicative correction factor of G.\ Brammer and I.\ Labbe presented in \citet{brammer2022a}, \num{0.7845}.
The correction factors are presented for the \textit{F090W}, \textit{F115W}, \textit{F150W}, and \textit{F200W}, that is, only the last band overlaps between the filter sets.
To remain consistent between all four NIRCam bands, we did not change the value of the apparent magnitude, but determine the uncertainty from the multiplicative correction factor itself.
The final uncertainty on the magnitude was then determined through Gaussian error propagation of this systematic uncertainty and the statistical uncertainty obtained from the fit.
For the other three NIRCam bands, we assumed that the correction factor equals \num{0.8}, resulting in an uncertainty of $\sim \SI{0.24}{\mag}$.
As for the \textit{F200W}, we combined this value with the statistical uncertainty from bootstrapping for the final uncertainty.

For the {\jwst} MIRI data, the background level was, as described in Section \ref{subsec:james_webb_space_telescope}, adjusted by comparing the \textit{F770W} flux to the \textit{IRAC4} \SI{8}{\micro\metre} and \textit{WISE3} \SI{12}{\micro\metre} bands.
The estimated uncertainty on its value is $\pm \SI{0.1}{\mag}$ \citep[][this issue]{LEROY1_PHANGSJWST}.

As pointed out in Section \ref{subsec:point_spread_functions}, we did not check the influence the calibration and co-addition of the data have on the synthetic PSF generated by {\webbpsf}.
\citet{hoyer2022a} found for the {\hst} data that the co-addition of single exposures results in a slight broadening of the core of the PSF, introducing a systematic overestimation of the NSCs size.
To repeat this experiment for the {\jwst} data, we focused on the NIRCam \textit{F200W} band and repeated the fit introducing an additional jitter in the form of a Gaussian function convolved with the synthetic PSF.
In {\webbpsf}, we increased the jitter by factors of two and five and repeated the fits using the new PSFs.
The result was that the structural parameters, as well as magnitude and color, remained well within the $\num{1} \sigma$ distribution of the original fits.
Therefore, we conclude that our results are reliable given the presented uncertainties.
\section{Analysis}
\label{sec:analysis}

\subsection{Photometry}
\label{subsec:photometry}

Integrating \Cref{equ:sersic} with an assumed ellipticity ($\epsilon$) yields the luminosity ($L$; photon count per energy band and time) as 
\begin{equation}
  \label{equ:sersic_integrated}
  L = \num{2} \pi \, (\num{1} - \epsilon) \, I_{\mathrm{eff}} \, r_{\mathrm{eff}}^{\num{2}} \, \frac{n \mathrm{e}^{b_{n}}}{(b_{n})^{\num{2}n}} \, \Gamma(\num{2}n) \; .
\end{equation}
We use the equation
\begin{align}
  \label{equ:abmag1}
  \mathrm{ZP}_{\mathrm{AB}} = & \num{-2.5} \, \log_{10} \, (\, \mathrm{PHOTFLAM} \,) \nonumber \\
                              & \num{-5} \, \log_{10} \, (\, \mathrm{PHOTPLAM} \,) \nonumber \\
                              & \num{-2.408} \; ,
\end{align}
to calculate the zero point magnitudes for {\hst} ACS / WFC3 bands.
The values of \textit{PHOTFLAM} and \textit{PHOTPLAM} are given in the header extensions of the fits files.%
\footnote{We find the following zero point magnitudes for the {\hst} bands: $\mathrm{ZP}_{\mathrm{\textit{F275W}}} = \SI{24.159}{\mag}$, $\mathit{ZP}_{\mathrm{\textit{F336W}}} = \SI{24.689}{\mag}$, $\mathit{ZP}_{\mathrm{\textit{F435W}}} = \SI{25.677}{\mag}$, $\mathit{ZP}_{\mathrm{\textit{F555W}}} = \SI{25.722}{\mag}$, $\mathrm{ZP}_{\mathrm{\textit{F658N}}} = \SI{22.760}{\mag}$, and $\mathrm{ZP}_{\mathrm{\textit{F814W}}} = \SI{25.950}{\mag}$.}

Pixel values in {\jwst} data products have the unit \si{\mega\jansky\per\stern}, which we convert to \si{\jansky} by using \Cref{equ:sersic_integrated} and the pixel-to-steradian conversion factor \textit{PIXAR\_SR} from the header extension.
The zero point magnitude is then derived using
\begin{equation}
    \label{equ:abmag2}
    \mathrm{ZP}_{\mathrm{AB}} = \num{-2.5} \, \log_{10} \, (L) + \num{8.9} \; .
\end{equation}

Foreground extinction is taken into account by using the re-calibrated version of the \citet{schlegel1998a} extinction maps \citep{schlafly2011a} and assuming $R_{V} = \num{3.1}$ \citep{fitzpatrick1999a}.
Due to the apparent lack of dust in the center of {\obj}, we do not attempt to correct for intrinsic extinction.
For the {\hst} bands, we derive $A(\lambda) \, / \, A(\mathrm{\textit{V}})$ with the model from \citet{odonnell1994a}, which is based on \citet{cardelli1989a}.

\begin{figure*}
  \centering
  \includegraphics[width=0.75\textwidth]{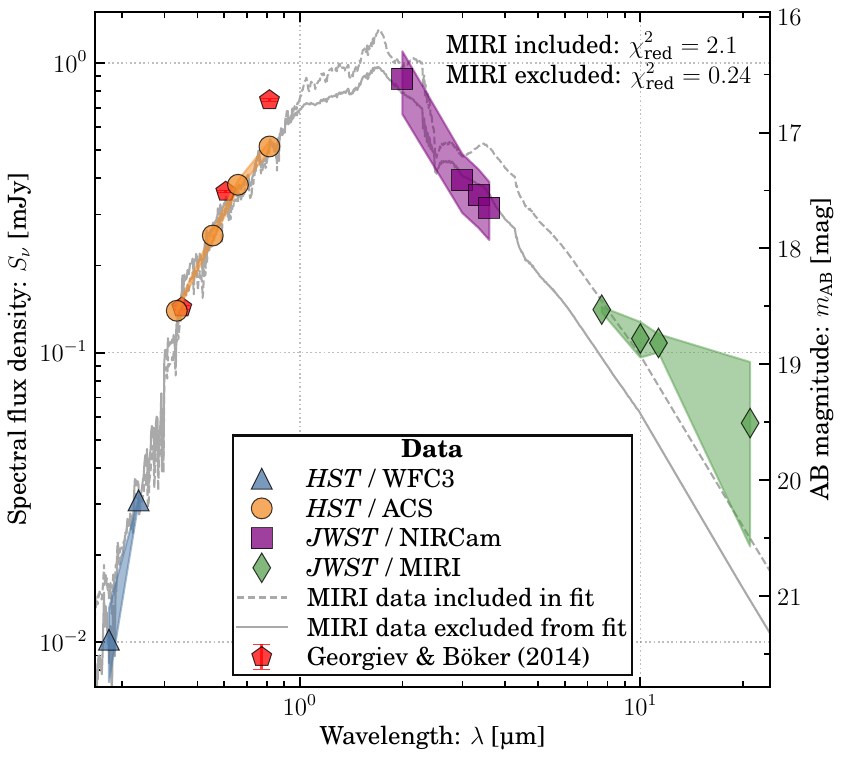}
  \caption{%
    Spectral flux density ($S_{\nu}$, \textit{left axis}) and AB magnitude ($m_{\mathrm{AB}}$, \textit{right axis}) versus wavelength ($\lambda$).
    Different instruments are highlighted with marker symbols and colors.
    Red pentagons give the results by \citet{georgiev2014a} in the Vega-magnitude system.
    Uncertainties are indicated with shaded areas.
    Two gray lines show the spectral energy distribution fits to the data including (dashed line) and excluding (solid line) the {\jwst} MIRI data.
    The goodness of the fits is given by reduced $\chi^{2}$ values in the panel.
    For both fits, we use the \citet{bruzual2003a} stellar population model and a delayed star formation history.
    The fit prefers a \citet{chabrier2003a} initial mass function over the prescription by \citet{salpeter1955a}.
    Both fits indicate the presence of a \SI{8}{\giga\yr} old stellar population with metallicity $Z \sim 0.02$.
    No younger stellar population could be detected.
  }
  \label{fig:sed_fit}
\end{figure*}

\Cref{fig:sed_fit} shows spectral flux densities as well as the extinction-corrected apparent magnitudes of the NSC in the AB-magnitude system.
The NSC is faintest in the ultraviolet regime and becomes brighter towards the near-infrared.
The brightest magnitude is reached at \SI{2}{\micro\metre} after which the nucleus becomes fainter again.

To compare to the values by \cta{georgiev2014a}, we transform our magnitudes from the AB- to the Vega-magnitude system using the approach outlined in \citet{sirianni2005a} and applied in \citet{pechetti2020a} for NSCs.
We find $\mathrm{\textit{V}}_{0} = (\num{17.85} \pm \num{0.04}) \, \si{\mag}$ and $\mathrm{\textit{I}}_{0} = (\num{16.69} \pm \num{0.05}) \, \si{\mag}$.
\cta{georgiev2014a} present $\mathrm{\textit{V}}_{0} = (\num{17.88} \pm \num{0.01}) \, \si{\mag}$ and $\mathrm{\textit{I}}_{0} = (\num{16.57} \pm \num{0.01}) \, \si{\mag}$.
While the $V$-band magnitudes agree with each other, we find a significant difference in the \textit{I}-band.
The different magnitude is most likely related to the extracted structural parameters (\textit{cf}.\ Section \ref{subsec:structure}).

\subsection{SED modeling}
\label{subsec:sed_fitting}

The combined {\hst} and {\jwst} data cover the ultraviolet to mid-infrared spectrum and enable the study of the spectral energy distribution (SED) in detail.
To extract basic parameters describing the stellar population, we set up a model assuming a delayed star formation history, two commonly used different initial mass functions \citep{chabrier2003a,salpeter1955a}, and the \citet{bruzual2003a} single stellar population model.

The fits were executed using \texttt{CIGALE}, a Python code for modeling the SEDs of galaxies \citep[see][and \citealp{burgarella2005a,noll2009a,yang2020a,yang2022a}]{boquien2019a}, which has successfully been applied to star clusters \citep[e.g.][]{fensch2019a,turner2021a}.
The program allows to adjust various physical properties such as the age of the stellar populations or their metallicity.
We test various parameter values, as detailed in \Cref{tab:sed_parameters}, to find the best possible fit to the data, as evaluated by Bayesian statistics.

\begin{deluxetable*}{
    l
    l
    l
    l
    l
    l
  }
  \tablecaption{%
    Parameters values for the spectral energy distribution fits.
    The values remain unchanged between runs including and excluding the MIRI data.
    \label{tab:sed_parameters}
  }
  \tablehead{
    \multicolumn{1}{c}{Parameter} & \multicolumn{1}{c}{Unit} & \multicolumn{1}{c}{Values} & \multicolumn{2}{c}{Best-fit} & \multicolumn{1}{c}{Notes}\\
                                  &                          &                            & Incl.\ MIRI & Excl.\ MIRI & 
  }
  \startdata
    \multicolumn{6}{c}{Star formation history} \\ \addlinespace
    {tau\_main}  & [\si{\mega\yr}] & \num{1}, \num{10}, \num{100}, \num{1000}, \num{2000}, \num{3000}                                      & $\num{500} \pm \num{500}$ & $\num{220}^{+380}_{-220}$ & e-folding time of the main stellar population \\
    {age\_main}  & [\si{\giga\yr}] & \num{3}, \num{4}, \num{5}, \num{6}, \num{7}, \num{8}, \num{9}, \num{10}, \num{11}, \num{12}, \num{13} & $\num{8} \pm \num{2}$     & $\num{8} \pm \num{3}$ & Age of the main stellar population \\
    {tau\_burst} & [\si{\mega\yr}] & \num{10}, \num{20}, \num{50}, \num{100}                                                               & $\num{57} \pm \num{33}$   & $\num{57} \pm \num{33}$ & Time of the late starburst \\
    {age\_burst} & [\si{\mega\yr}] & \num{10}, \num{20}, \num{50}, \num{100}                                                               & $\num{27} \pm \num{17}$   & $\num{27} \pm \num{17}$ & Age of the late burst \\
    {f\_burst}   & -{-}            & \num{0.0}, \num{0.1}, \num{0.2}, \num{0.5}                                                            & $\num{0} \pm \num{0}$     & $\num{0} \pm \num{0}$ & Mass fraction of the late burst \\ \addlinespace[10pt]
    %
    \multicolumn{6}{c}{Simple stellar population} \\ \addlinespace
    {imf}         & -{-} & \num{0}, \num{1}                                 & $\num{1}$                     & $\num{1}$ & Initial mass function (\citealp{salpeter1955a}, \citealp{chabrier2003a}) \\
    {metallicity} & -{-} & \num{0.004}, \num{0.008}, \num{0.02}, \num{0.05} & $\num{0.027} \pm \num{0.013}$ & $\num{0.012} \pm \num{0.006}$ & Metallicity of the stellar population
  \enddata
\end{deluxetable*}

The fit was performed twice, excluding the MIRI data in one run.
We do this to test their influence to the fit result and disentangle emission from low-mass stars and dust.
The addition of a dust emission model yielded worse fits, as evaluated by both the reduced $\chi^{2}$ and Bayesian statistics, which is why we do not include it in the presented results.
We discuss this issue in more detail in Section \ref{sec:discussion}.

For the fit including the MIRI data, we find that the mass-weighted age of the main stellar population is $(\num{8} \pm \num{2}) \, \si{\giga\yr}$ with a metallicity of $Z = \num{0.03} \pm \num{0.01}$, slightly more metal-rich than the Sun \citep{asplund2009a}.
The e-folding time of the main stellar population is $(\num{500} \pm \num{500}) \, \si{\mega\yr}$ and the mass fraction of the late burst is consistent with zero.
The fit preferred a \citet{chabrier2003a}-type initial mass function over a \citet{salpeter1955a} one.

For the fit excluding the MIRI data, we find a mass-weighted age of the main stellar population of $(\num{8} \pm \num{3}) \, \si{\giga\yr}$ with a metallicity of $Z = \num{0.012} \pm \num{0.006}$.
The e-folding time was determined to be $\num{220}_{\num{-220}}^{\num{+380}} \, \si{\mega\yr}$ and the mass fraction of the late burst is comparable to zero.
The fit again preferred a \citet{chabrier2003a}-type initial mass function.

According to the reduced $\chi^{2}$ statistics, the fit excluding the MIRI data performed better than the one including them.
The results of both fits are consistent with each other, indicating the presence of a \SI{8}{\giga\yr} old stellar population with metallicity $Z \sim \num{0.02}$ and no presence of a young stellar population.
We discuss the results obtained for the mass of the stellar population in the next section.

\subsection{Stellar mass}
\label{subsec:stellar_mass}

We determine the stellar mass of the NSC in three different ways:
(1) we use $\mathrm{\textit{B}} - \mathrm{\textit{V}}$ color and its mass-to-light scaling relations, (2) we combine the apparent magnitude in the \textit{F200W} (roughly \textit{K}-band) with a constant mass-to-light ratio ranging between \num{0.5} and \num{0.6} (in solar units), and (3) we extrapolate a stellar mass from SED fitting.
The resulting mass estimates and the literature value from \citet{georgiev2016a} are presented in \Cref{tab:nscmass}.

Following \citet{hoyer2021a}, we use four different $\mathrm{\textit{B}} - \mathrm{\textit{V}}$ color relations and the \textit{V}-band luminosity to obtain a stellar mass-to-light ratio.
The original relations were published by \citet{bell2003a,portinari2004a,zibetti2009a,into2013a} but we adopt the revised parameters from \citet{mcgaugh2014a}, which ensures consistency between the relations.
An extended discussion and the assumptions made are detailed in \citet{hoyer2021a}.

First, the {\hst} ACS \textit{F435W} and \textit{F555W} magnitudes were converted to the Johnson-Cousin system (\textit{B}- and \textit{V}-band, respectively) using Equation~\num{12} and Table~\num{22} from \citet{sirianni2005a}.
Absolute magnitudes were derived using the distance estimate of the galaxy and the absolute magnitude of the Sun \citep{willmer2018a}.%
\footnote{See \url{http://mips.as.arizona.edu/~cnaw/sun.html} for an overview. The uncertainty on the values is assumed to be \SI{0.04}{\mag}.}

After transforming the {\hst} magnitudes to the Johnson-Cousin system and using magnitudes in the Vega-system, we find $\mathrm{\textit{B}}_{0} = (\num{18.59} \pm \num{0.03}) \, \si{\mag}$ and $\mathrm{\textit{V}}_{0} = (\num{17.86} \pm \num{0.04}) \, \si{\mag}$ with a color $(\mathrm{\textit{B}} - \mathrm{\textit{V}})_{0} = (\num{0.73} \pm \num{0.05}) \, \si{\mag}$, which roughly matches the color of a \texttt{G8V}-type star ($\mathrm{\textit{B}} - \mathrm{\textit{V}} \sim 0.75$) and is consistent with the results from the SED fits.
From the four scaling relations, we determine individual masses and combine them into one using the weighted average.
The resulting mass estimate is $\log_{10} \, (M_{\star}^{\mathrm{nsc}} \, / \, \si{\Msun}) = \num{7.06} \pm \num{0.31}$.
The uncertainty budget is dominated by the uncertainty assumed for the mass-to-light ratio, \SI{0.3}{\dex} \citep{roediger2015a}.

An alternative approach is to use the magnitude in the \textit{K}-band.
\citet{mcgaugh2014a} found that a constant mass-to-light ratio of $\sim \num{0.6}$ can be used to estimate stellar masses as the near-infrared luminosity is only weakly dependent on color.
While we do not directly have a \textit{K}-band magnitude, we estimate the mass using the \textit{F200W} band from {\jwst}, centered on \SI{2}{\micro\metre}.
The \textit{K}-band overlaps with the \textit{F200W} band such that we can use the mass estimate as a benchmark for the other mass estimates.

Using the same four references as for the approach using the $\mathrm{\textit{B}} - \mathrm{\textit{V}}$ color \citep{bell2003a,portinari2004a,zibetti2009a,into2013a} and their re-calibrated values from \citet{mcgaugh2014a}, we find a stellar mass of $\log_{10} \, (M_{\star}^{\mathrm{nsc}} \, / \, \si{\Msun}) = \num{7.2} \pm \num{1.1}$.
The uncertainty is much larger than for the mass determined from the $\mathrm{\textit{B}} - \mathrm{\textit{V}}$ relation due to the uncertainty on the zero point of the NIRCam data.

From the SED fitting in the previous section, the mass of the star cluster was determined as well.
In the fit including the MIRI data, we find $\log_{10} \, (M_{\star, \, 1}^{\mathrm{nsc}} \, / \, \si{\Msun}) = \num{7.17} \pm \num{0.10}$.
In the fit excluding the MIRI data, we find $\log_{10} \, (M_{\star, \, 2}^{\mathrm{nsc}} \, / \, \si{\Msun}) = \num{7.11} \pm \num{0.10}$.
As stated above, no young stellar population was found.

The mass of the NSC was previously determined by \citet{georgiev2016a} based on the analysis of \cta{georgiev2014a}.
To obtain stellar masses, the authors use stellar population models from \citet{bruzual2003a} with solar metallicity and an initial mass function of the type presented in \citet{kroupa2001a}.
The reported mass for the NSC of {\obj} is $\log_{10} \, (M_{\star}^{\mathrm{nsc}} \, / \, \si{\Msun}) = \num{7.05} \pm \num{0.21}$, which agrees within the uncertainty with our mass estimates.

Overall, we find agreement between all approaches finding that the NSC has a stellar mass of $\sim \SI{e7}{\Msun}$.
In the following, we use the mass value $\log_{10} \, (M_{\star}^{\mathrm{nsc}} \, / \, \si{\Msun}) = \num{7.06} \pm \num{0.31}$.
\begin{deluxetable}{
    l
    l
    l
    l
  }
  \tablecaption{%
    Determined nuclear star cluster mass as well as the literature value.
    \label{tab:nscmass}
  }
  \tablehead{
    \multicolumn{1}{c}{Method} & \multicolumn{1}{c}{Logarithmic stellar mass} \\
     & \multicolumn{1}{c}{[\si{\Msun}]}
  }
  \startdata
    $\mathrm{\textit{B}} - \mathrm{\textit{V}}$ & $\num{7.06} \pm \num{0.31}$ \\
    \textit{K}-band                             & $\num{7.2} \pm \num{1.1}$\tablenotemark{{\small a}} \\
    SED (incl.\ MIRI)                           & $\num{7.17} \pm \num{0.10}$ \\
    SED (excl.\ MIRI)                           & $\num{7.11} \pm \num{0.10}$ \\ 
    \citet{georgiev2016a}                       & $\num{7.05} \pm \num{0.21}$ 
  \enddata
  \tablenotetext{{\small \textit{a}}}{The large uncertainty compared to the other values is caused by the high uncertainty on the zero point values for NIRCam data.} 
\end{deluxetable}

\subsection{Structure}
\label{subsec:structure}

\begin{figure*}
  \centering
  \includegraphics[width=\textwidth]{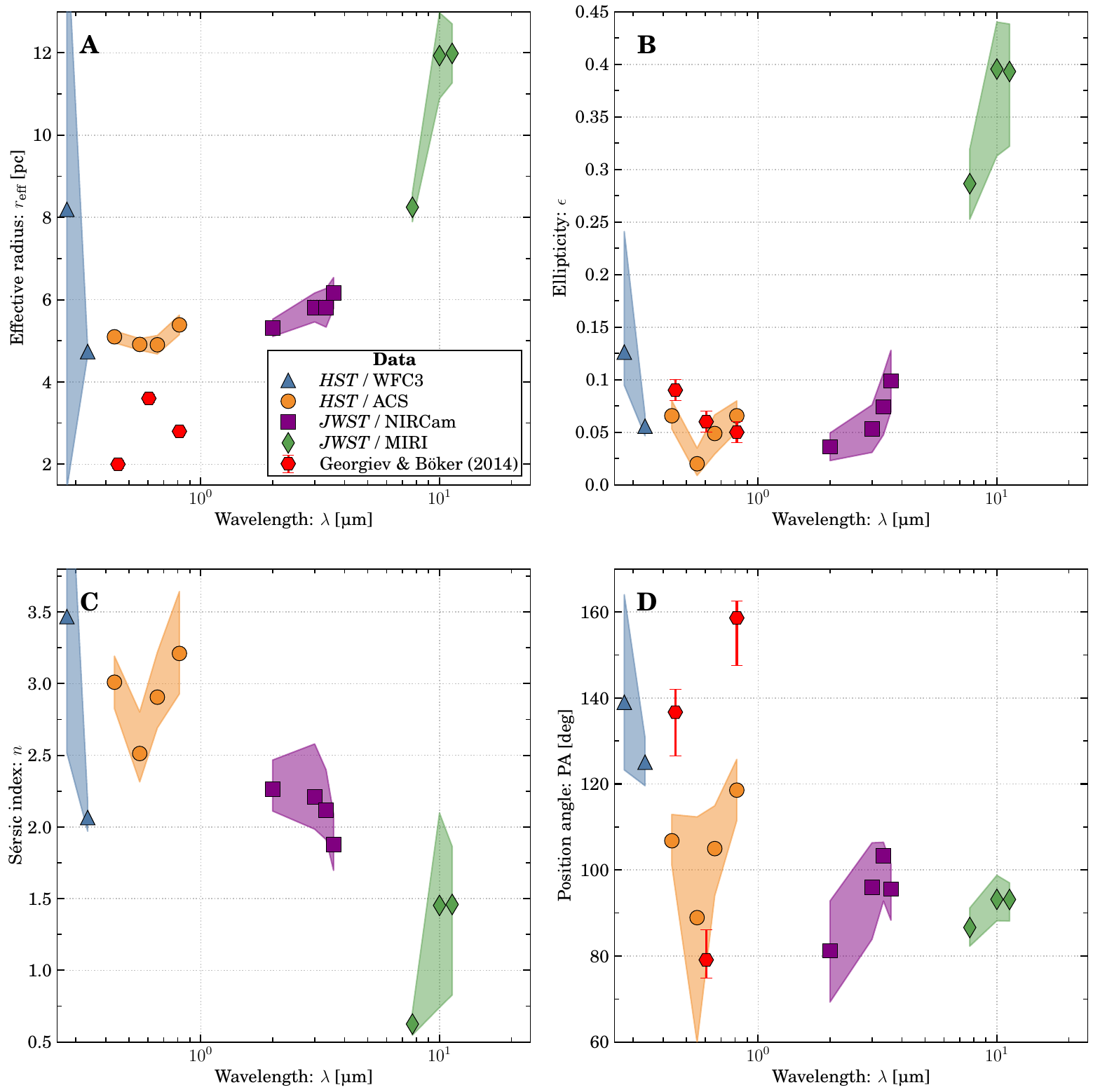}
  \caption{%
    Structural properties of the nucleus of {\obj}:
    effective radius (panel A), ellipticity (panel B), S{\'{e}}rsic index (panel C), and position angle (panel D) versus wavelength.
    Marker symbols and colors highlight different instruments.
    Uncertainties are shown with shaded areas.
    Red pentagons show literature values from \citet{georgiev2014a}.
  }
  \label{fig:structure}
\end{figure*}

\Cref{fig:structure} shows the effective radius, ellipticity, S{\'{e}}rsic index, and position angle versus wavelength (from Section \ref{sec:image_fitting}).
We also add the literature values by \cta{georgiev2014a}.

In panel A, we show the effective radius versus wavelength.
It remains roughly constant at $\sim \SI{5}{\pc}$ in the ultraviolet and optical regime, but starts to slightly increase towards $\sim \SI{6}{\pc}$ at \SI{3.6}{\micro\metre}.
This trend continues into the mid-infrared where $r_{\mathrm{eff}} \sim \SI{12}{\pc}$.

\cta{georgiev2014a} find different effective radii ranging from $\sim \SI{2}{\pc}$ to $\sim \SI{3.5}{\pc}$.
They modeled the NSC light distribution by convolving a {\tinytim}-generated PSF with King profiles of different concentration parameters (ratio of the tidal to core radius: \num{5}, \num{15}, \num{30}, and \num{100}).
Using \texttt{ISHAPE} \citep{larsen1999a}, they fit the data and used the best-fit model, according to $\chi^{2}$ residuals to derive the effective radius of the cluster.
In their fits, a concentration parameter of \num{100} gave the best results.
The final value for the effective radius was obtained by taking the geometric mean of the full-width-half-maximum along the semi-minor and major axes and using a transformation factor from \texttt{ISHAPE}'s manual.

In the ultraviolet and optical regime, the ellipticity is $\sim \num{0.05}$ (panel B in \Cref{fig:structure}).
It remains in this range at \SI{2}{\micro\metre} and \SI{3}{\micro\metre}, but starts to increase to $\sim \num{0.1}$ at \SI{3.6}{\micro\metre}.
At even longer wavelengths, the ellipticity increases to $\sim \num{0.4}$ and is significantly different from the other wavelength regimes.
Our measurements in the optical are consistent with the ones presented by \cta{georgiev2014a}, but are smaller than the typical ellipticity for other NSCs in the same mass range \citep[$\epsilon \gtrsim \num{0.1}$, e.g.][]{seth2006a,carson2015a,spengler2017a,hoyer2022a}.

The S{\'{e}}rsic index (panel C) appears to vary with wavelength.
In the ultraviolet and optical regime, we find $n \sim \num{3}$, but in the near-infrared the value drops to $\sim \num{2}$.
At the longest wavelengths, the value drops to $\sim \num{1.5}$, but is also consistent with an exponential profile ($n = \num{1}$).
\cta{georgiev2014a} used a King profile to approximate the light distribution and no comparison can be made.

The position angle of the NSC (panel D) starts at $\sim \SI{130}{\degree}$ in the ultraviolet regime.
Starting in the optical regime, the position angle drops to $\sim \SI{100}{\degree}$ and shows a mild anti-correlation with wavelength, dropping further to $\sim \SI{90}{\degree}$ in the mid-infrared.
Only the data point from the {\hst} WFPC2 PC \textit{F606W} band by \cta{georgiev2014a} is consistent with our results.
The other two data points are significantly elevated to $\sim \SI{135}{\degree}$ and $\sim \SI{160}{\degree}$.

\subsection{Astrometric offset}
\label{subsec:astrometric_offset}

From the previous section it is apparent that the nucleus shows an evolution with wavelength, especially towards the mid-infrared regime.
Here we investigate the variability of the central position of the emission in different bands.

\begin{figure}
  \centering
  \includegraphics[width=\columnwidth]{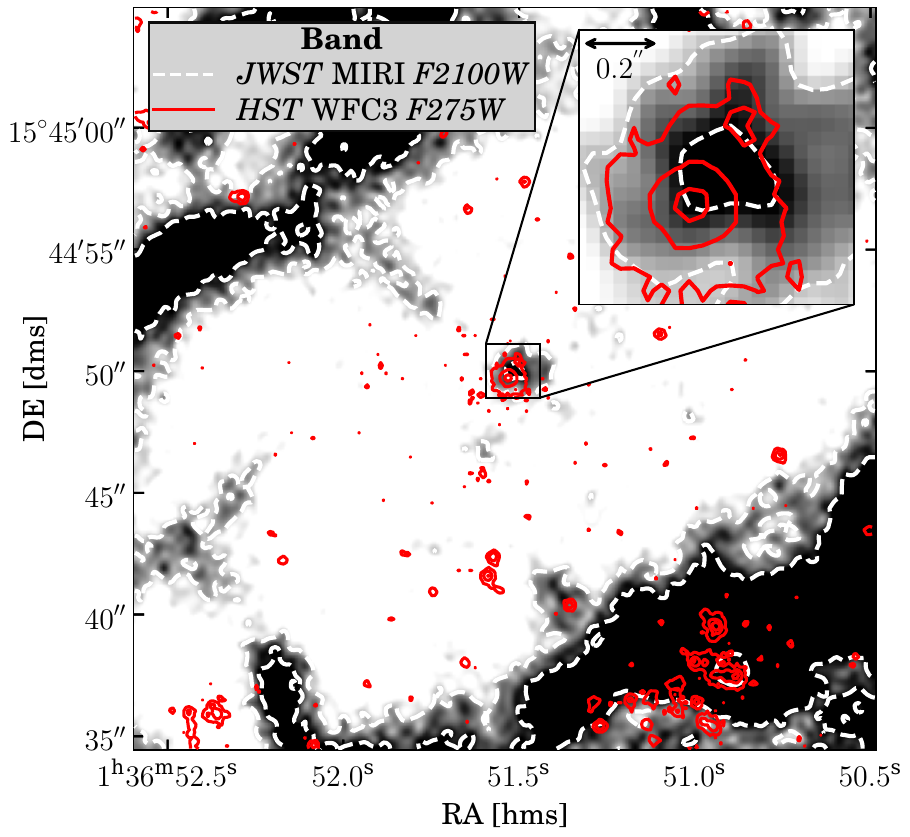}
  \caption{%
    \textit{Main panel}:
    Central $\SI{5.5}{\arcsecond} \times \SI{5.5}{\arcsecond}$ of {\obj} in the {\jwst} MIRI \textit{F2100W}.
    A darker shade and white contours show high flux in the \textit{F2100W}.
    Red contours highlight the {\hst} WFC3 \textit{F275W} emission.
    Both maps are matched based on the most up-to-date world coordinate solutions.
    \textit{Inset panel}:
    Zoom into the nucleus of {\obj}.
    The white and red contours highlight again the emission in the \textit{F2100W} and \textit{F275W}, respectively.
    The offset between the centers of the contours measures approximately \SI{0.2}{\arcsecond}.
  }
  \label{fig:offset1}
\end{figure}

In \Cref{fig:offset1} we show the emission in the {\jwst} MIRI \textit{F2100W} band (gray scale and white contour lines) and overlay the emission from the {\hst} WFC3 \textit{F275W} band.
The WCS of each band were taken from the bands header files.
We find that there exists an offset between the centers of the emission, separated by $\sim \SI{0.2}{\arcsec}$, which approximates to $\sim \SI{9.5}{\pc}$ at the distance to the galaxy.

\begin{figure}
  \centering
  \includegraphics[width=\columnwidth]{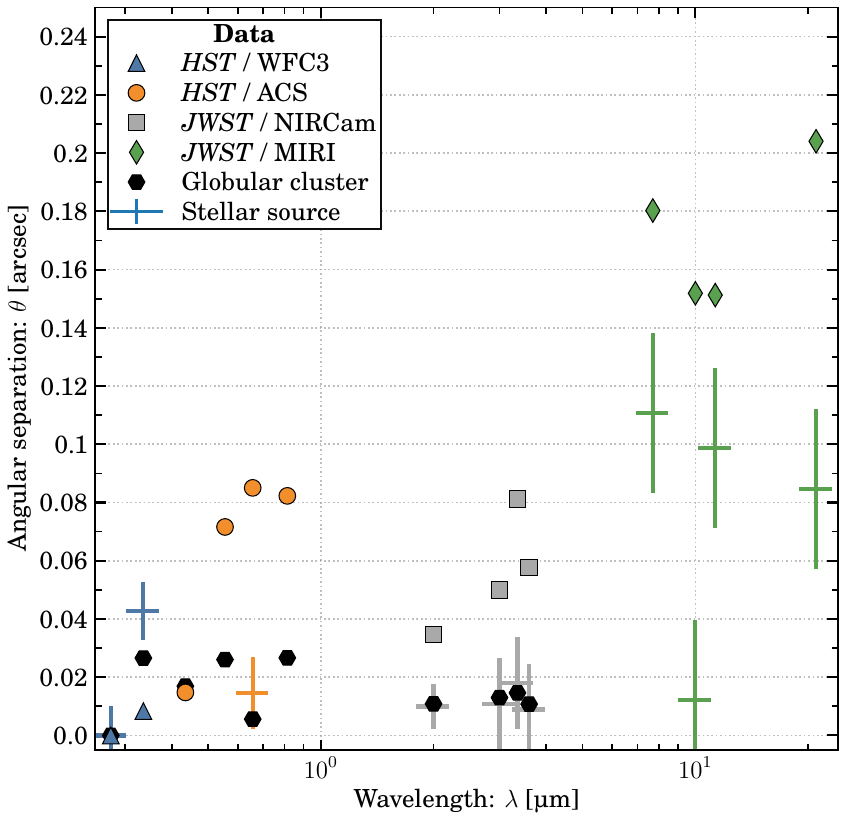}
  \caption{%
    Angular separation ($\theta$) versus wavelength ($\lambda$) of the nuclear star cluster position with respect to the {\hst} WFC3 \textit{F275W}.
    Different instruments are highlighted with different marker symbols and colors.
    Crosses show the angular separation of a star with \textit{Gaia} EDR3 designation \texttt{2589386446469602688}.
    The vertical length of the cross give \SI{0.5}{\pixel}, which we assume as an upper limit on extracting the position of the star.
    Data points above the crosses indicate that either the world coordinate system is offset, or the center of the nucleus is offset compared to the {\hst} WFC3 \textit{F275W}.
    In some bands (for example the {\hst} ACS \textit{F814W}), the star is saturated and no central position could be determined.
    The separations for a globular cluster, located $\sim \SI{10}{\arcsec}$ South-West of the nucleus is shown with black hexagons.
    The offsets agree with the offsets of the star in the NIRCam data.
  }
  \label{fig:offset2}
\end{figure}

To test whether the offset persists in other bands, we perform the following experiment:
we determine the angular coordinates for other bands based on the central position of the S{\'{e}}rsic profile fit to the light distribution.
\Cref{fig:offset2} presents the resulting angular separation using the {\hst} WFC3 \textit{F275W} band as reference.
We find that the angular separation is of the order of $\lesssim \SI{0.1}{\arcsecond}$ ($\lesssim \SI{5}{\pc}$) in the optical, which is comparable to the effective radius of the cluster.
In the infrared the separation drops to $\lesssim \SI{0.06}{\arcsecond}$ but increases up to \SI{0.21}{\arcsecond} in the mid-infrared regime.

Depending on the band used as reference, the angular separation can become insignificant.
For example, while using the \textit{F200W} as reference, the offsets in the MIRI bands are still significant, whereas they become insignificant, except for the \textit{F1000W}, when using the \textit{F335M} as reference.
This behavior could point towards issues with the calibration of the WCS':
while we calibrated all {\hst} data with the most recent WCS solutions from MAST, no reliable solutions exist so far for the {\jwst} data.
The PHANGS-internal versions of the data were calibrated in the following way:
The NIRCam data was calibrated using {\hst} and \textit{Gaia} astrometric solutions using asymptotic giant branch stars.
Furthermore, the direction of the separation is the same in the MIRI bands, towards the North-West (as seen in \Cref{fig:offset1}).

Compared to the {\hst} data, the NIRCam calibration should yield ``accurate'' astrometric values (see below).
The MIRI data are astrometrically aligned to the \textit{F335M} image by cross-correlating the images and solving for relative offsets.
However, due to variations in the polycyclic aromatic hydrocarbons emission structure between different bands and the lack of point-like emission in the MIRI bands, the astrometric calibration is less certain.

To further quantify the offsets and benchmark the values, we compute the angular separation of (a) a star outside the central cavity, (b) multiple stars less than \SI{1}{\arcsec} South of the NSC within the cavity, and (c) a GC about \SI{10}{\arcsec} South-West from the NSC.
The star outside the cavity has \textit{Gaia} EDR3 designation \texttt{2589386446469602688}, is non-saturated in all but the {\hst} ACS \textit{F435W}, \textit{F555W}, and \textit{F814W} bands and lies about \SI{60}{\arcsecond} South-East of the NSC.
For the star outside the cavity and the GC, we fit the light distribution with a two-dimensional Gaussian function, which yields the position of the center of the sources, which we deem accurate within \SI{0.5}{\pixel}.
For the other stars close to the NSC, we extract the position manually.

In \Cref{fig:offset2} we show the offset of the star outside the cavity and the GC in addition to the offsets for the NSC.
If we use the {\hst} WFC3 \textit{F275W} data as reference, the offsets are significant in both the NIRCam and MIRI data.
The same result is found if we use the \textit{F200W} as reference.
However, the offsets become insignificant in all except two bands (\textit{F200W} \& \textit{F1000W}) if we use the \textit{F335M} data as reference.
With the currently available WCS calibrations, while there are hints of an astrometric offset, we cannot conclude whether they are significant.

\section{Discussion}
\label{sec:discussion}

One of the most striking features of the {\jwst} observations of the center of {\obj} is that the prominent NSC sits in a nuclear stellar component that is devoid of gas and dust.
It appears that both gas and dust have been evacuated from the central cavity.
The mechanism that created this cavity is not obvious.
There are no young stars that could have blown out the gas recently.
Alternatively, the central cavity could have been created by consumption of the gas in the last star formation event, and the re-supply of gas is  hindered by a potential bar resonance, in case a bar is (or previously was) present.
\footnote{As mentioned in the introduction, \citet{querejeta2021a} find that {\obj} hosts no bar.}

\subsection{Nuclear star cluster properties}
\label{subsec:nuclear_star_cluster_properties}

\begin{figure*}
  \centering
  \includegraphics[width=\textwidth]{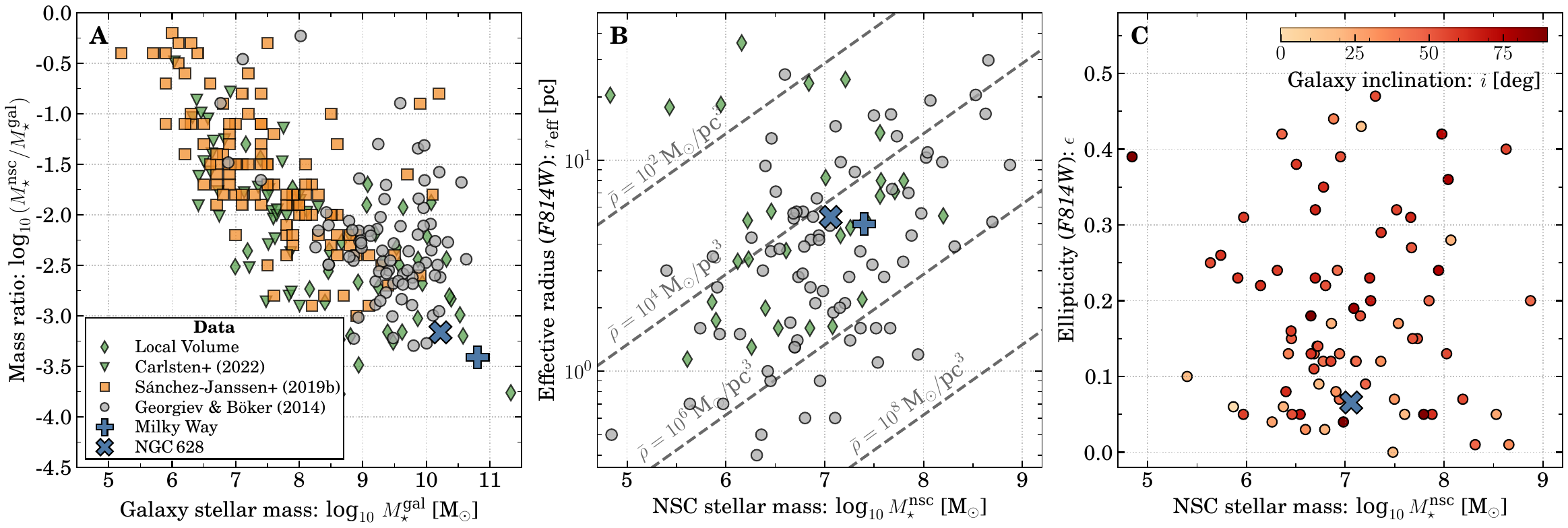}
  \caption{%
    \textit{Left panel}:
    Mass ratio of nuclear star cluster (NSC) and host galaxy stellar mass ($M_{\star}^{\mathrm{nsc}} / M_{\star}^{\mathrm{gal}}$) versus galaxy stellar mass.
    The NSC of {\obj} is highlighted with a blue cross.
    Other data sets from the Local Volume (a field environment with a distance $\lesssim \SI{11}{\mega\pc}$; green diamonds; \citealp{seth2006a,georgiev2009b,graham2009b,baldassare2014a,schoedel2014a,calzetti2015a,carson2015a,crnojevic2016a,nguyen2017a,baumgardt2018b,nguyen2018a,bellazzini2020a,pechetti2020a}), the Virgo galaxy cluster (orange squares; \citealp{sanchez-janssen2019a}), dwarfs around massive field galaxies (green triangles; \citealp{carlsten2022a}), and massive field galaxies (gray circles; \citealp{georgiev2014a}) show the general distribution of NSCs.
    The Milky Way NSC is highlighted with a blue plus-sign.
    In comparison to other massive field late-type galaxies, the NSC of {\obj} appears under-massive.
    \textit{Middle panel}:
    Effective radius ($r_{\mathrm{eff}}$) in the \textit{F814W} versus NSC stellar mass.
    The markers and colors of the data points are the same as in the left panel.
    Dashed lines give the mean density of clusters.
    \textit{Right panel}:
    Ellipticity ($\epsilon$) versus NSC stellar mass.
    We show the data by \citet{georgiev2014a} and color-code them by the host galaxy inclination.
    There exists no apparent correlation between ellipticity, mass, or inclination.
    For {\obj}, the inclination has a value of $i \sim \SI{8.9}{\degree}$ \citep{lang2020a}.
  }
  \label{fig:comparison}
\end{figure*}

We show the mass ratio (NSC mass divided by host galaxy mass) in the left panel of \Cref{fig:comparison} where the NSC of {\obj} is highlighted with a blue cross.
Data from the Local Volume (a field environment with distance $\lesssim \SI{11}{\mega\pc}$; \citealp{seth2006a,georgiev2009b,graham2009b,baldassare2014a,schoedel2014a,calzetti2015a,carson2015a,crnojevic2016a,nguyen2017a,baumgardt2018b,nguyen2018a,bellazzini2020a,pechetti2020a}) are added for comparison.
Dwarfs around massive field galaxies and Virgo cluster members are taken from \citet{carlsten2022a} and \citet{sanchez-janssen2019a}, respectively.
\footnote{Although not considered here, data for nucleated dwarf galaxies in the Fornax galaxy cluster is presented by \citet{munoz2015a,eigenthaler2018a,venhola2018a,ordenes-briceno2018b,su2021a}.}
Data for other massive late-type galaxies in the field are taken from \cta{georgiev2014a}.
The NSC of {\obj} follows the overall trend in that the NSC mass becomes insignificant compared to the host galaxy.
However, other late-types of the same host galaxy mass typically host more massive NSCs.

The effective radius of the NSC in {\obj} also compares well to those of other NSCs in late-type galaxies (in the \textit{F814W} band; middle panel in \Cref{fig:comparison}).
Finally, the ellipticity in the \textit{F814W} band is smaller than the typical value in other late-types (right panel).
This figure shows there exists no apparent correlation with the inclination of the host.

Since the mass of the NSC is smaller than most other masses of such a cluster at $M_{\star}^{\mathrm{gal}} \sim \SI{e10}{\Msun}$ and assuming that stars formed \textit{in-situ} should dominate the mass budget, we speculate that the NSC had a quiet evolution and that, compared to other NSCs, only little mass formed \textit{in-situ} over the last few \si{\giga\yr}.
This speculation is corroborated by our results in Sections \ref{subsec:photometry}, \ref{subsec:sed_fitting}, and \ref{subsec:structure}:
the effective radius shows no wavelength dependence from the ultraviolet to the near-infrared regime, staying roughly constant at \SI{5}{\pc}.
The color of the NSC was determined to be $\mathrm{\textit{B}} - \mathrm{\textit{V}} = (\num{0.73} \pm \num{0.05}) \, \si{\mag}$, which compares to a star of \texttt{G8V}-class. 
Finally, the resulting best-fit SED model indicate that all stellar mass is assembled in an ``old'' stellar population, with the mass of the ``young'' stellar population being consistent with zero.
As indicated by the fit, ``old'' refers to an age of \SI{8}{\giga\yr}.
If true, this could also mean that the cavity has existed for a few \si{\giga\yr} and that any massive black hole in the center of {\obj} did not grow significantly via gas accretion over the same time period.
So far, no reliable black hole mass measurement is available (see also Section \ref{subsubsec:agn_contribution} below).

The SED fit also indicates that the metallicity of the NSC is $Z \sim \num{0.02}$, which is comparable to NSCs in similar mass galaxies \citep{koleva2009a,paudel2011a,spengler2017a,kacharov2018a,neumayer2020a}, and also the Milky Way NSC \citep{do2015a,feldmeier-krause2017a}.
In combination with the  age estimate of the stellar population, this reveals that the NSC formed in a dense environment where rapid enrichment took place.
Such conditions could take place either during the formation of the galaxy itself or during a past merger event.

As mentioned above, while \textit{in-situ} star formation is expected to contribute a significant mass fraction to the NSC, as measured in other galaxies, our results indicate that no \textit{in-situ} star formation occurred over the last few \si{\giga\yr}.
This means that, since the formation of the NSC, either no or very little amount of gas fell towards the center or that star formation was inefficient.
One possibility is that the shape of the gravitational potential limits the amount of inflow.
Indeed, it is well known that in a viscous accretion disk, the amount of inward mass transport depends on the amount of shear \citep[e.g.][]{shakura1973a,lynden-bell1974a}.
One way to limit the inflow of gas is to have a low shear, meaning that the rotation curve is close to solid body rotation \citep[e.g.][]{lesch1990a,krumholz2015e}.
Note, however, that it is still unclear what mechanism drives the ISM turbulence responsible for creating the required viscosity \citep[e.g.][]{klessen2016a,sormani2020a}.
Alternatively, it could be that the in-flow is irregular and triggered by mergers or interactions with satellite galaxies \citep[e.g.][]{storchi-bergmann2019a}.
Multiple dwarfs are known to reside around {\obj} \citep{davis2021a} and numerous accretion events occurred in the galaxy's history \citep{kamphuis1992a}.

\subsection{Comparison with the Milky Way}

The obtained NSC size using near-infrared data seems to be very similar to the Milky Way’s (MWNSC) with an effective radius of $\sim \SI{5}{\pc}$ \citep[e.g.][]{fritz2016a,gallego-cano2020a}.
However, the MWNSC also presents a similar size when analyzed with Spitzer/IRAC mid-infrared data \citep{schoedel2014b,gallego-cano2020a}, which is in contrast to the significantly larger effective radius we obtained for the NSC of {\obj} from MIRI mid-infrared data.
In addition, the mass estimates compare, with the MWNSC having a mass of $\sim \SI{2e7}{\Msun}$ \citep{launhardt2002a,schoedel2014a,feldmeier-krause2017b}.

The predominantly old ($\sim \SI{8}{\giga\yr}$) and metal-rich ($Z \sim \num{0.02}$) population detected in {\obj}'s NSC is also in agreement with the results obtained for the MWNSC \citep[e.g.][]{feldmeier-krause2017a,schoedel2020a,nogueras-lara2022b}, though recent work suggested a younger age for the MWNSC of $\sim \SI{5}{\giga\yr}$ \citep{chen2022c}.
However, the MWNSC also shows recent star formation activity, about \SI{6}{\mega\yr} ago \citep{paumard2006a}, which is not present in {\obj}'s NSC, according to the best-fit SED model.


Overall, we find that little to no star formation occurred in the last few \si{\giga\yr} in {\obj}'s center.
This results in an under-massive NSC, compared to other similar-mass late-type galaxies, a likely under-massive central black hole, if present, and that the central cavity spanning approximately $\SI{200}{\pc} \times \SI{400}{\pc}$ existed for a similar period.

\subsection{Nature of the emission in the mid-infrared}
\label{subsec:nature_of_the_emission_in_the_mid-infrared}

While an old (\SI{8}{\giga\yr}) population with metallicity $Z \sim \num{0.02}$ accounts for the emission in the ultraviolet to near-infrared regime, we found an excess of emission in the mid-infrared bands (\textit{cf}.\ \Cref{fig:sed_fit}), which cannot be explained by that same population.
In addition, the effective radius and ellipticity do not change with wavelength until the mid-infrared regime, the S{\'{e}}rsic index shows a weak wavelength dependence, and the position angle does not change significantly between the near- and mid-infrared (\textit{cf}.\ \Cref{fig:structure}).
We speculate about the nature of the emission in the following sections.

\subsubsection{Active galactic nucleus contribution}
\label{subsubsec:agn_contribution}

One possibility is that the emission in the mid-infrared bands is caused by an active galactic nucleus (AGN) once X-ray photons are absorbed by dust, which re-emits the radiation at longer wavelengths.

The presence of a massive black hole in {\obj} is still disputed:
\citet{dong2006a} use the black hole mass versus bulge $\mathrm{\textit{K}}_{\mathrm{\textit{S}}}$-magnitude relation to find $\log_{10} \, (M_{\mathrm{BH}} \, / \, \si{\Msun}) \sim \num{6.7}$ but such a relation assumes that the bulge did not significantly grow through secular processes, which is believed to be the case for {\obj}.
\citet{she2017a} found an X-ray excess in the galaxy's center, which they attribute to the presence of an AGN with a black hole mass of $\log_{10} \, (M_{\mathrm{BH}} \, / \, \si{\Msun}) \sim \num{6.0}$.
The X-ray luminosity was determined to be $\log_{10} \, (L_{2\textrm{-}\SI{10}{\kilo\eV}} \, / \, \si{\watt}) = \num{31.15}^{+\num{0.32}}_{\num{-0.19}}$, as determined through the hardness ratios of soft-, medium-, and hard X-ray bands.

We determine the spectral flux density of the emission in the mid-infrared by using this luminosity and the scaling relation by \citet{asmus2015a}, which connects the X-ray luminosity of an AGN to the mid-infrared luminosity at \SI{12}{\micro\metre}.
The result is $S_{\nu}^{\mathrm{BH}} \sim \SI{1.9e-4}{\milli\jansky}$.
We compare this value to the difference between the observed emission and the model flux excluding the MIRI data in the \SI{11.3}{\micro\metre} band.
The difference equals $\Delta S_{\nu}^{\SI{11.3}{\micro\metre}} \sim \SI{0.06}{\milli\jansky}$, far exceeding the expected flux density of an AGN.
Therefore, while the X-ray excess measured by \citet{she2017a} originating from a possible AGN could contribute to the mid-infrared emission, it cannot fully explain it by itself.
Furthermore, little to no dust is present in the NSC, making this scenario unlikely.

\subsubsection{Infalling star cluster}
\label{subsubsec:an_in-falling_star_cluster}

A possible scenario, which could perhaps explain the offset in \Cref{fig:offset1}, if real, is that we see the NSC and an in-falling star cluster, where the latter could be in a late stage of tidal disruption by the more massive NSC.
Such a scenario for the build-up of NSCs has been proposed for a few decades \citep{tremaine1975a} and is sometimes referred to as the ``dry-merger'' scenario \citep[e.g.][]{arca-sedda2018a} with ample observational and theoretical evidence in both the Galactic but also extragalactic NSCs 
\citep[e.g.][]{antonini2013a,antonini2014b,arca-sedda2017b,fahrion2020a,feldmeier-krause2020a}.

The proposed scenario could occur as follows:
the star cluster would form outside the nuclear region and spiral inwards.
During this time, the star cluster can be considered self-gravitating, which implies that it evolved predominantly due to its internal collisional dynamics.
During the infall of the cluster, it will experience gravothermal-gravogyro contraction and core-collapse \citep[e.g.][]{kamlah2022b}, mass segregate, and form a subsystem of black holes in its center, or even an intermediate-mass black hole, if the star cluster is massive enough.
The most-massive stars accumulate in the star cluster's center and lower-mass stars occupy the halo of the star cluster.
Some of these low-mass stars will be stripped by the tidal field of the surrounding field or may be ejected through dynamical interactions, while the star cluster approaches the NSC.
Some of the stripped or ejected stars might be visible as asymptotic giant branch (AGB) stars (see also Section \ref{subsubsec:dust_from_agb_stars}) with their strong, dust-driven stellar winds \citep[see][and sources therein]{decin2021a} in the near- to mid-infrared bands as single sources scattered around the NSC (see \Cref{fig:fit_jwst}).

From $N$-body simulations by \citet{arca-sedda2018a}, modeling the MWNSC and an infalling star cluster, we know what the infall, merger, and merger product phases look like in spatial coordinates (Figure 2 in \citealp{arca-sedda2018a} and Figure 1 in \citealp{arca-sedda2020c}).
If the infalling star cluster has already crossed the effective radius of the NSC, after which the star cluster becomes entirely tidally disrupted and cannot be considered a self-gravitating system anymore \citep{arca-sedda2020c}, the simulation snapshots could explain the potential astrometric offset.
The star cluster's core would eventually fall into the core of the NSC and the remaining halo stars would tidally disperse.
Among these would be AGB stars that may partly be responsible for the astrometric offset shown in \Cref{fig:offset1} and contribute to the elliptical increase in panel B of \Cref{fig:structure} (see also Section \ref{subsubsec:dust_from_agb_stars} below).

One counter-argument is that it is unlikely to witness such an event:
\citet{arca-sedda2020a} simulated the infall of a star cluster on an NSC whose properties mimic the ones of the Milky Way NSC.
They find that the star clusters enters a region \SI{10}{\pc} around the center of the NSC after \SI{60}{\mega\yr} and that the cluster is not a self-gravitating system anymore after another \SI{1}{\mega\yr}.
Note that the bulge component in their simulation is likely more massive than the bulge-component of {\obj} and that the time scale for in-spiral will be longer.
Nevertheless, the time scale will be short compared to the age of the cluster, $\sim \SI{8}{\giga\yr}$.

\subsubsection{Dust from AGB stars}
\label{subsubsec:dust_from_agb_stars}

While on the AGB, the outer layers of a star expand drastically leading to a circum-stellar envelope, which leads to an enrichment of the interstellar medium, contributing to the mass budget for future star formation \citep[e.g.][]{loup1997a,van_loon1998a}.
Material from the stellar winds can produce dust, which cools off and becomes visible in the mid-infrared regime.
Note that the dust would reside ``close'' to the star (at a few hundred stellar radii for a temperature of $\sim \SI{100}{\kelvin}$; \citealp{decin2021a}), thus not obscuring the emission of other stars in the NSC, which is why we observe no dust obscuration in the ultraviolet and optical regime.
Here we explore whether AGB stars can account for the emission in the MIRI bands (\textit{cf}.\ \Cref{fig:sed_fit}).

We first determine the residual flux, which is not accounted for by the SED model excluding the data.
The residual values are $\Delta S_{\nu} =$ \num{0.096}, \num{0.062}, \num{0.049}, and \SI{0.014}{\milli\jansky} in the \textit{F770W}, \textit{F1000W}, \textit{F1130W}, \textit{F2100W}, respectively.
Next, we generate absolute magnitudes of AGB stars using \texttt{PARSEC} tracks \citep{bressan2012a}, with \SI{60}{\percent} Silicate and \SI{40}{\percent} AlOx for M-type stars, and \SI{85}{\percent} AMC and \SI{15}{\percent} SiC for C-type stars \citep{groenewegen2006a}, long-period variabilities from \citet{trabucchi2021a}, a log-normal \citet{chabrier2003a} initial mass function, and a metallicity of $Z = \num{0.012}$.%
\footnote{The models were calculated by \url{http://stev.oapd.inaf.it/cgi-bin/cmd_3.6}, \citealp{bressan2012a,chen2014a,chen2015a,tang2014a,marigo2017a,pastorelli2019a,pastorelli2020a}.}
The last two settings equal the results found from SED fitting.
We then convert the absolute magnitudes to spectral flux densities using the distance estimate to {\obj} and Vega- to AB-magnitude conversion factors for the Sun \citep{willmer2018a}.

\begin{figure}
  \centering
  \includegraphics[width=\columnwidth]{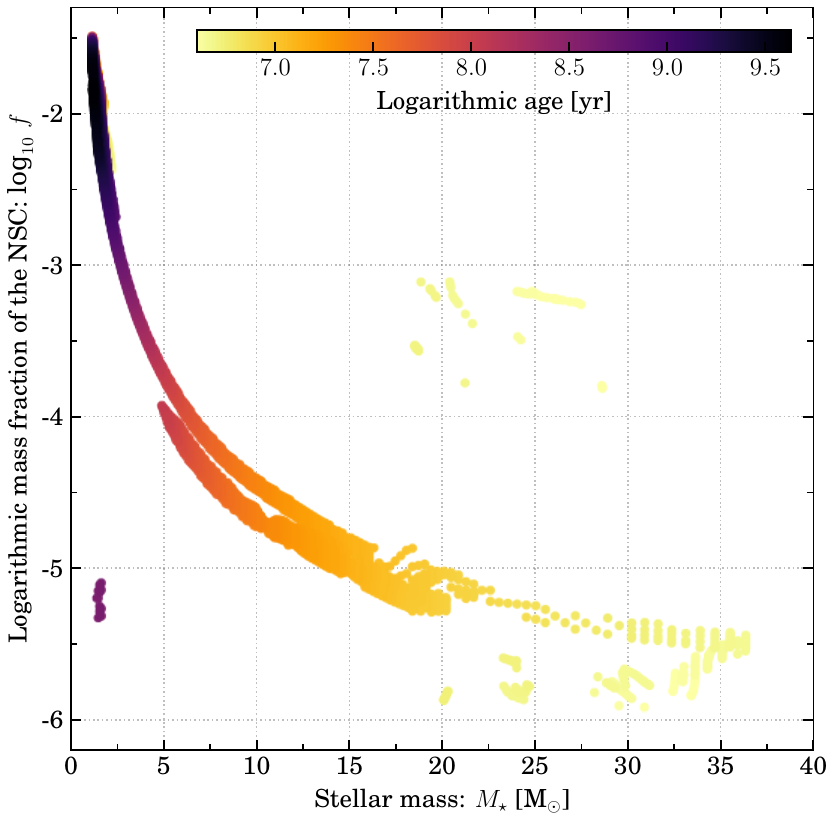}
  \caption{%
    Logarithmic fraction of the number of AGB stars multiplied by their mass and divided by the total nuclear star cluster mass ($\log_{10} \, f$) versus AGB star mass.
    Each data point is color-coded according to the age of the star.
    Note that all AGB stars are younger than the main stellar population of the NSC, as identified by fitting the spectral energy distribution.
  }
  \label{fig:agb_stars}
\end{figure}

First, we limit the AGB model stars to reside within the $1 \sigma$ interval of the measured colors.%
\footnote{We use the six colors $\mathrm{\textit{F770W}}-\mathrm{\textit{F1000W}}$, $\mathrm{\textit{F770W}}-\mathrm{\textit{F1130W}}$, $\mathrm{\textit{F770W}}-\mathrm{\textit{F2100W}}$, $\mathrm{\textit{F1000W}}-\mathrm{\textit{F1130W}}$, $\mathrm{\textit{F1000W}}-\mathrm{\textit{F2100W}}$, and $\mathrm{\textit{F1130W}}-\mathrm{\textit{F2100W}}$.}
Afterwards, we determine how many AGB stars are required to account for the residual emission in the MIRI bands and multiply that number by the mass of the stars.
\Cref{fig:agb_stars} shows the logarithmic mass fraction of AGB stars compared to the total NSC mass versus the mass of the individual AGB stars.
The data are color-coded by the age of the AGB stars.
We find that both a few young and many old AGB stars could be responsible for the emission in the mid-infrared.

However, all model AGB stars that satisfy the color-cuts are younger than \SI{5}{\giga\yr}, which gives the lower uncertainty on the age of the main stellar population of the NSC.
Therefore, if AGB stars are responsible for the emission in the MIRI data, there must have been star formation \textit{in-situ} after the initial formation of the NSC.

In case the AGB stars are old, meaning that many AGB stars are required to account for the emission in the mid-infrared, it remains unclear why both the effective radius and ellipticity change significantly, as the cluster with the AGB star should have relaxed between their formation and today.
In contrast, only few massive and young AGB stars are required to account for the mid-infrared emission, which could explain the increased effective radius and ellipticity, if they formed outside the center of the NSC.
However, this would require in-flow of gas in the past few \si{\mega\yr} but we cannot detect the presence of a young stellar population in the NSC.
Therefore, it remains challenging to explain both the structural and photometric parameters using only AGB stars.

\subsubsection{A circum-nuclear gaseous disk}
\label{subsubsec:circumnuclear_gas_disk}

Another possibility is that the infrared emission originates from a circum-nuclear gaseous disk or ring with a radius of a few pc, similar to the one present in the MW.
Indeed, the MW hosts a clumpy, asymmetric, inhomogeneous, and kinematically disturbed concentration of molecular/ionized gas at $R \lesssim \SI{5}{\pc}$ known as the circum-nuclear disk \citep[e.g.][]{lau2013a,hsieh2021a}.
The MW circum-nuclear disk occupies similar radii to its NSC, has a total mass of $M_{\mathrm{gas}}^{\mathrm{disk}} \simeq \num{e4}$-$\SI{e5}{\Msun}$ and it is probably a transient structure (on a timescale of few \si{\mega\yr}) originating from a series of randomly oriented in-flow events \citep{requena-torres2012a}.
By analogy, we could hypothezise that {\obj} hosts a similar gaseous structure and that this is producing the observed mid-infrared emission.
While the emission from the gas disk could explain the observed photometry, it is unclear why we do not detect a ``young'' (formed in the last few \si{\giga\yr}) stellar population in the NSC.
While the \textit{ALMA} CO band does not show significant emission in {\obj}'s center (see \Cref{fig:overview}), this may be related to the sensitivity of the ALMA measurements and could not exclude a low-mass disk:
in the the PHANGS--ALMA v4p0 ``broad'' CO~(2-1) map, the intensity measurement at the position of the NSC is $I_{(\mathrm{CO}) 2-1} = (-0.27 \pm \num{1.30}) \si{\kelvin\kilo\metre\per\second}$. Given the beam size ($\theta=\SI{1.12}{\arcsec}$) and distance to the target, translating this value to a $\num{3} \sigma$ upper limit yields to the CO~(2-1) luminosity yields $\log_{10} \, L_{(\mathrm{CO}) 2-1} \, / \, \si{\kelvin\kilo\metre\per\second\pc\squared} < 4.1$. 
For a standard Milky Way CO to H$_{2}$ conversion factor and CO~(2--1) to CO~(1--0) line ratio appropriate for NGC~628 \citep{bolatto2013b,denbrok2021a}, this luminosity limit corresponds to an upper mass limit of $\log_{10} \, M_{\mathrm{H}_{2}} \, / \, \si{\Msun} < 4.9$.
In comparison, the circum-nuclear disk of the Milky Way has a mass of $M_{\mathrm{gas}} \sim \SI{1.2e4}{\Msun}$ \citep{requena-torres2012a}.

\subsubsection{Background galaxy}
\label{subsubsec:background_galaxy}

It is also plausible that the emission in the MIRI data originates from a background galaxy, which happens to be aligned with the NSC along the line of sight.
Although an alignment of the order \SI{0.1}{\arcsecond} is unlikely, we investigate this scenario further based on the photometry found in the MIRI data.

\citet[][this issue]{HASSANI_PHANGSJWST} investigate the properties of compact sources at \SI{21}{\micro\metre} for all four PHANGS--JWST targets for which data are available.
Using a dendogram-based algorithm, they find \num{1271} compact sources of which \num{115} are classified as ``potential background sources'' (or HZ).
This classification was performed using flux density ratios between MIRI bands (their Equations 1 and 2).
The MIRI structure coinciding with the NSC of {\obj} was also classified as a potential background object.

A search in the NASA Extragalactic Database\footnote{\url{https://ned.ipac.caltech.edu/}} revealed that the \num{114} sources were previously detected by the \textit{WISE} / \textit{ALLWISE} mission \citep{wright2010a,cutri2013a} and all objects were classified as ``infrared sources''.
While these sources show a galaxy-like morphology at \SI{2}{\micro\metre}, their detailed properties remain unclear at this point.

To compare to the other potential background objects, we select the measured spectral flux densities for the MIRI bands and subtract the extrapolated emission from the NSC using the SED fit excluding the MIRI data (solid line in \Cref{fig:sed_fit}).
While the flux density values compare to other potential background sources, their evolution with wavelength does not:
none of the \num{114} identified potential background objects follow a similar trend in that the flux densities decrease with increasing wavelength.

Therefore, if the other \num{114} sources are background galaxies, the differences in the evolution of flux densities with wavelength suggest that the MIRI emission coinciding with the NSC of {\obj} is not related to a background galaxy.
Such a scenario becomes more unlikely if we combine it with the probability of alignment with the NSC along the line of sight.
\section{Conclusions}
\label{sec:conclusions}

In this work we analysed the nuclear star cluster (NSC) of {\obj}, a nearby late-type spiral galaxy, with archival \textit{Hubble Space Telescope} ({\hst}) ACS \& WFC3 and newly obtained \textit{James Webb Space Telescope} ({\jwst}) NIRCam \& MIRI data.
The combined data cover the ultraviolet to mid-infrared wavelength, enabling an unprecedented analysis of an extragalactic NSC.
Our findings can be summarized as follows:
\begin{enumerate}
  \item
  Combining the $\mathrm{\textit{B}} - \mathrm{\textit{V}}$ color with various mass-to-light relations results in an NSC stellar mass of $\log_{10} \, (M_{\star}^{\mathrm{nsc}} \, / \, \si{\Msun}) = \num{7.06} \pm \num{0.31}$.
  We compare this number to an estimate derived using the \textit{K}-band magnitude (resulting in $\num{7.2} \pm \num{1.1}$) and the results from fitting the spectral energy distribution (SED; resulting in $\sim \num{7.1}$).
  These values are consistent with the literature value of $\num{7.05} \pm \num{0.21}$ \citep{georgiev2016a}.

  \item
  The effective radius and ellipticity of the NSC are $\sim \SI{5}{\pc}$ and $\sim \num{0.05}$, respectively, across the ultraviolet, optical, and near-infrared regime.
  At the same time the S{\'{e}}rsic index drops from $\sim \num{3}$ to $\sim \num{2}$ and the position angle drops from $\sim \SI{130}{\degree}$ to $\sim \num{90}$-$\SI{100}{\degree}$.
  These values supersede literature values, which varied significantly across neighboring bands \citep{georgiev2014a}. 

  \item
  In the mid-infrared bands, the effective radius and ellipticity increase to $\sim \SI{12}{\pc}$ and $\sim \num{0.4}$, respectively.
  The S{\'{e}}rsic index drops further to $\sim \num{1.5}$ while being consistent with an exponential profile, and the position angle remains unchanged compared to the near-infrared.

  \item
  We fit the SED from the ultraviolet to the near-infrared with a total of ten data points to find an old stellar population of $(\num{8} \pm \num{3}) \si{\giga\yr}$ with a metallicity of $Z = \num{0.012} \pm \num{0.006}$.
  The fit indicates that no younger stellar population is present.

  \item
  Fitting the SED with the inclusion of the MIRI data yields an overall worse fit, as evaluated by $\chi^{2}$ statistics.
  Nevertheless, the age and metallicity of the main stellar population remain unchanged within the uncertainties.
  The differences in both fits to the SED indicate that the MIRI data do not trace the stellar population found in the lower wavelength regimes.

  \item
  We find different angular separations between the center of the NSC in different bands, being most significant in the mid-infrared data.
  However, depending on the band from which the world coordinate system is taken as reference, the separations become less significant.
  This could hint at persistent calibration issues with the world coordinate systems of individual bands.
\end{enumerate}

The color, age, and metallicity of the main stellar population of {\obj}'s NSC indicate that no star formation has taken place in the previous few \si{\giga\yr} in its center.
This is caused either by a dynamical mechanism preventing gas and dust inflow, or by feedback from the center.
The lack of a young stellar population hints that the central cavity, which lacks both gas and dust and has a size of approximately $\SI{200}{\pc} \times \SI{400}{\pc}$ around the NSC, has existed for the last few \si{\giga\yr} as well.
The reason for the lack of recent \textit{in-situ} star formation and origin of the central cavity remains unknown.

The nature of the emission in the mid-infrared bands remains a mystery as well.
From our SED fits it is clear that the old stellar population of the NSC cannot completely explain the emission in {\jwst}'s MIRI bands.
We discussed five different mechanisms, which may cause the emission:
(1) contribution from a central active galactic nucleus, (2) an infalling star cluster, (3) dust from asymptotic giant branch (AGB) stars, (4) the presence of a circum-nuclear disk, and (5) alignment with a background galaxy.

The AGB scenario could explain the observed photometry.
However, we find that the AGB stars whose colors fit the measurements are younger than the main stellar population.
From a comparison to model AGB stars, we find that either a large number of old or a small number of young stars are required.
While the first scenario cannot explain the wavelength dependence of the structural parameters, the latter scenario requires recent (a few \si{\mega\yr} ago) in-falling gas, however, no stellar population younger than \SI{5}{\giga\yr} was detected from SED fitting.
In conclusion, none of the four discussed scenarios can fully explain both the structural and photometric measurements.

Our analysis highlights the potential {\jwst} data has for exploring galactic nuclei in the nearby Universe.
An ongoing analysis of the stellar population using PHANGS--MUSE data can improve the situation, albeit it cannot resolve the NSC.
To solve the riddle of the nucleus of {\obj} at long wavelengths, we will propose high-resolution spectroscopic observations, ideally Integral Field Unit spectroscopic data with {\jwst}, to determine the kinematic properties of the NSC and its direct surroundings.
In addition to the nature of the structure in the mid-infrared bands, these data will constrain further the presence of a young stellar population, the kinematic signature compared to the cluster, and help to constrain the presence of a black hole in {\obj} as there is currently no available robust mass measurement or upper limit.

\section*{Acknowledgments}
This research is based on observations made with the NASA/ESA Hubble Space Telescope obtained from the Space Telescope Science Institute, which is operated by the Association of Universities for Research in Astronomy, Inc., under NASA contract  NAS 5–26555. These observations are associated with program 15654.
This work is based on observations made with the NASA/ESA/CSA JWST. The data were obtained from the Mikulski Archive for Space Telescopes at the Space Telescope Science Institute, which is operated by the Association of Universities for Research in Astronomy, Inc., under NASA contract NAS 5-03127. The observations are associated with JWST program 02107.
%
This research has made use of the Spanish Virtual Observatory (\url{https://svo.cab.inta-csic.es}) project funded by MCIN/AEI/10.13039/501100011033/ through grant PID2020-112949GB-I00.

NH and AWHK are fellows of the International Max Planck Research School for Astronomy and Cosmic Physics at the University of Heidelberg (IMPRS-HD) and acknowledge their support.
NH acknowledges support from Thomas M{\"{u}}ller (HdA/MPIA) for support with generating part of  \Cref{fig:overview}, and Katja Fahrion and Torsten B{\"{o}}ker for useful discussions.
ATB and FB would like to acknowledge funding from the European Research Council (ERC) under the European Union’s Horizon 2020 research and innovation programme (grant agreement No.726384/Empire).
EJW acknowledges the funding provided by the Deutsche Forschungsgemeinschaft (DFG, German Research Foundation) -- Project-ID 138713538 -- SFB 881 (``The Milky Way System'', subproject P1)
TGW  and JN acknowledge funding from the European Research Council (ERC) under the European Union’s Horizon 2020 research and innovation programme (grant agreement No. 694343).
JMDK gratefully acknowledges funding from the European Research Council (ERC) under the European Union's Horizon 2020 research and innovation programme via the ERC Starting Grant MUSTANG (grant agreement number 714907). 
COOL Research DAO is a Decentralized Autonomous Organization supporting research in astrophysics aimed at uncovering our cosmic origins.
RSK acknowledges funding from the European Research Council via the ERC Synergy Grant ``ECOGAL'' (project ID 855130), from the Deutsche Forschungsgemeinschaft (DFG) via the Collaborative Research Center ``The Milky Way System''  (SFB 881 -- funding ID 138713538 -- subprojects A1, B1, B2 and B8) and from the Heidelberg Cluster of Excellence (EXC 2181 - 390900948) ``STRUCTURES'', funded by the German Excellence Strategy. RSK also thanks the German Ministry for Economic Affairs and Climate Action for funding in the project ``MAINN'' (funding ID 50OO2206).
ER acknowledges the support of the Natural Sciences and Engineering Research Council of Canada (NSERC), funding reference number RGPIN-2022-03499.
MB acknowledges support from FONDECYT regular grant 1211000 and by the ANID BASAL project FB210003.
KG is supported by the Australian Research Council through the Discovery Early Career Researcher Award (DECRA) Fellowship DE220100766 funded by the Australian Government. 
KG is supported by the Australian Research Council Centre of Excellence for All Sky Astrophysics in 3 Dimensions (ASTRO~3D), through project number CE170100013.
F. N.-L. gratefully acknowledges the sponsorship provided by the Federal Ministry for Education and Research of Germany through the Alexander von Humboldt Foundation.
PSB acknowledges financial support from the MCIN/AEI/10.13039/501100011033 under the grant  PID2019-107427GB-C31
AKL gratefully acknowledges support by grants 1653300 and 2205628 from the National Science Foundation, by award JWST-GO-02107.009-A, and by a Humboldt Research Award from the Alexander von Humboldt Foundation.
G.A.B. acknowledges the support from ANID Basal project FB210003.

%

\vspace{5mm}
\facilities{%
  HST(STIS), JWST(STIS)
}


\software{%
  Astropy \citep{astropy2013a,astropy2018a},
  AplPy \citep{robitaille2012a,robitaille2019a},
  dustmaps \citep{green2018a},
  Imfit \citep{erwin2015a},
  Matplotlib \citep{hunter2007a},
  NumPy \citep{harris2020a},
  Photutils \citep{bradley2020a}
  TinyTim \citep{krist1993a,krist1995a},
  WebbPSF \citep{perrin2012a,perrin2014a}
}

\section*{Data Availability}
The data underlying this article are publicly available at the Milkulski Archive for Space Telescopes (MAST)\footnote{\url{https://archive.stsci.edu/}}.
The specific observations analyzed can be accessed via \dataset[doi:10.17909/t9-r08f-dq31]{http://dx.doi.org/10.17909/t9-r08f-dq31} and \dataset[doi:10.17909/9bdf-jn24]{http://dx.doi.org/10.17909/9bdf-jn24}.



\appendix

\section{Number of S{\'{e}}rsic profiles}
\label{sec:number_of_sersic_profiles}

The description of the projected light distribution of NSCs is often approximated by a single simple analytic function such as a S{\'{e}}rsic profile, with few exceptions \citep{nguyen2018a,pechetti2022a}.
With increasing spatial resolution at all wavelength ranges, accurate descriptions of the light distribution of NSCs may warrant multiple profiles.
The NSC of {\obj} was analyzed previously by \citet{georgiev2014a} and modeled with a single King profile, but the goodness of the fit was not indicated.

Here we explore whether adding a second S{\'{e}}rsic profile improves the fit compared to a single S{\'{e}}rsic profile.
The goodness of the two fits may not be compared via the standard $\chi^{2}$ statistics, as different number of free parameters are at play.
To compensate for the increased number of free parameters $k$, a penalty is introduced by adding a term linear in $k$ to the standard $\chi^{2}$ evaluation.
We use the Bayesian Information Criteria \citep[BIC,][]{schwarz1978a}, defined as
\begin{equation}
  \mathrm{BIC} = -2 \ln \, \mathcal{L} + k \ln \, N \; ,
\end{equation}
where $\mathcal{L}$ is the likelihood value and $N$ the total number of data points.
Model ($A$) is generally preferred over model ($B$) if $\Delta \mathrm{BIC} = \mathrm{BIC}_{B} - \mathrm{BIC}_{A} > 0$, but note that the BIC is a heuristic approach and includes approximations.

We highlight the results for the single (labeled ``A'') and double S{\'{e}}rsic profile (labeled ``B'') fits for the NSC in \Cref{tab:bic_comparison}.
The $F2100W$ is excluded from the list as the fit with two profiles for the NSC did not converge under any circumstance.

The conclusion from this experiment is that a single S{\'{e}}rsic profile is preferred over fitting two S{\'{e}}rsic profiles for the NSC.

\begin{deluxetable}{
    l
    l
    l
    l
    l
  }
  \tablecaption{%
    Differences in the Bayesian Information Criteria ($\Delta \mathrm{BIC}$) between two ($A$) and a single S{\'{e}}rsic profile ($B$) for the NSC of {\obj}.
    \label{tab:bic_comparison}
  }
  \tablehead{%
    \multicolumn{1}{c}{Band} & \multicolumn{1}{c}{$\mathrm{BIC}_{(A)}$} & \multicolumn{1}{c}{$\mathrm{BIC}_{(B)}$} & \multicolumn{1}{c}{$\Delta \mathrm{BIC}$} & \multicolumn{1}{c}{$\frac{\mathrm{BIC}_{(A)}}{\mathrm{BIC}_{(B)}}$}
  }
  \startdata
    $F275W$  & \num{269}   & \num{203}   & \num{66} & \num{1.33} \\
    $F336W$  & \num{225}   & \num{156}   & \num{69} & \num{1.44} \\
    $F435W$  & \num{234}   & \num{155}   & \num{79} & \num{1.51} \\
    $F555W$  & \num{227}   & \num{162}   & \num{65} & \num{1.40} \\
    $F658N$  & \num{216}   & \num{150}   & \num{66} & \num{1.44} \\
    $F814W$  & \num{284}   & \num{222}   & \num{62} & \num{1.28} \\ \addlinespace
    $F200W$  & \num{14191} & \num{14121} & \num{70} & \num{1.00} \\
    $F300M$  & \num{940}   & \num{870}   & \num{70} & \num{1.08} \\
    $F335M$  & \num{1167}  & \num{1047}  & \num{120} & \num{1.11} \\
    $F360M$  & \num{1133}  & \num{1105}  & \num{27} & \num{1.03} \\ \addlinespace
    $F770W$  & \num{421}   & \num{343}   & \num{78} & \num{1.23} \\
    $F1000W$ & DNF\tablenotemark{a} & \num{382} & -{-} & -{-} \\
    $F1130W$ & DNF\tablenotemark{a} & \num{482} & -{-} & -{-} \\
    $F2100W$ & DNF\tablenotemark{a} & DNF\tablenotemark{a} & -{-} & -{-} \\
  \enddata
  \tablenotetext{a}{%
    The fit failed to terminate or parameter values ran into boundary conditions in all attempts.
  }
\end{deluxetable}
\section{Data table}
\label{sec:data_table}

In \Cref{tab:parameters_one} we present the best-fit parameters using a single S{\'{e}}rsic profile.

\begin{deluxetable*}{
    l
    l
    l
    l
    l
    l
    l
    l
    l
  }
  \tablecaption{%
    Best-fit parameter estimates for the nuclear star cluster of {\obj} using a single S{\'{e}}rsic profile.
    The parameter values and their uncertainties were determined via \num{500} bootstrap iterations, where the best-fit value gives the median and the uncertainties the $1\sigma$ interval.
    \label{tab:parameters_one}
  }
  \tablehead{
    \multicolumn{1}{c}{Band} & \multicolumn{1}{c}{RA} & \multicolumn{1}{c}{DEC} & \multicolumn{1}{c}{PA} & \multicolumn{1}{c}{$\epsilon$} & \multicolumn{1}{c}{$n$} & \multicolumn{2}{c}{$r_{\mathrm{eff}}$} & \multicolumn{1}{c}{$m_{0}$\tablenotemark{{\small a}}} \\
    & \multicolumn{1}{c}{[hms]} & \multicolumn{1}{c}{[dms]} & \multicolumn{1}{c}{[deg]} & & & \multicolumn{1}{c}{[\si{\asec}]} & \multicolumn{1}{c}{[\si{\pc}]\tablenotemark{{\small b}}} & \multicolumn{1}{c}{[\si{\mag}]}
  }
  \startdata
    \textit{F275W}  & 01:36:41.742 & +15:47:01.167 & $139^{+16}_{-25}$     & $0.13^{+0.03}_{-0.12}$    & $3.5^{+1.0}_{-1.5}$    & $0.17^{+0.03}_{-0.14}$    & $8.2^{+6.7}_{-6.7}$     & $21.38^{+0.31}_{-0.16}$ \\
    \textit{F336W}  & 01:36:41.742 & +15:47:01.173 & $125.1^{+5.5}_{-5.7}$ & $0.056^{+0.009}_{-0.010}$ & $2.07^{+0.10}_{-0.11}$ & $0.099^{+0.002}_{-0.002}$ & $4.74^{+0.09}_{-0.09}$  & $20.18^{+0.02}_{-0.02}$ \\ \addlinespace
    \textit{F435W}  & 01:36:41.743 & +15:47:01.174 & $106.8^{+5.7}_{-6.1}$ & $0.066^{+0.013}_{-0.014}$ & $3.01^{+0.19}_{-0.18}$ & $0.107^{+0.003}_{-0.003}$ & $5.10^{+0.14}_{-0.14}$  & $18.54^{+0.02}_{-0.02}$ \\
    \textit{F555W}  & 01:36:41.738 & +15:47:01.216 & $89^{+29}_{-23}$      & $0.020^{+0.011}_{-0.015}$ & $2.51^{+0.20}_{-0.29}$ & $0.103^{+0.003}_{-0.003}$ & $4.91^{+0.14}_{-0.14}$  & $17.89^{+0.04}_{-0.02}$ \\
    \textit{F658N}  & 01:36:41.736 & +15:47:01.186 & $105^{+11}_{-10}$     & $0.049^{+0.020}_{-0.018}$ & $2.91^{+0.21}_{-0.31}$ & $0.103^{+0.003}_{-0.005}$ & $4.91^{+0.22}_{-0.22}$  & $17.45^{+0.04}_{-0.02}$ \\
    \textit{F814W}  & 01:36:41.738 & +15:47:01.227 & $118.6^{+7.1}_{-7.2}$ & $0.066^{+0.015}_{-0.014}$ & $3.21^{+0.28}_{-0.43}$ & $0.113^{+0.004}_{-0.005}$ & $5.39^{+0.24}_{-0.24}$  & $17.12^{+0.04}_{-0.03}$ \\ \addlinespace
    \textit{F200W}  & 01:36:41.741 & +15:47:01.133 & $81^{+12}_{-12}$      & $0.036^{+0.013}_{-0.013}$ & $2.26^{+0.15}_{-0.02}$ & $0.111^{+0.004}_{-0.004}$ & $5.31^{+0.21}_{-0.21}$  & $16.54^{+0.27}_{-0.27}$ \\
    \textit{F300M}  & 01:36:41.738 & +15:47:01.177 & $96^{+12}_{-10}$      & $0.053^{+0.022}_{-0.023}$ & $2.21^{+0.23}_{-0.37}$ & $0.122^{+0.006}_{-0.007}$ & $5.81^{+0.35}_{-0.35}$  & $17.41^{+0.25}_{-0.25}$ \\
    \textit{F335M}  & 01:36:41.737 & +15:47:01.219 & $103^{+11}_{-3}$      & $0.074^{+0.027}_{-0.032}$ & $2.12^{+0.21}_{-0.28}$ & $0.122^{+0.008}_{-0.010}$ & $5.80^{+0.47}_{-0.47}$  & $17.54^{+0.25}_{-0.25}$ \\
    \textit{F360M}  & 01:36:41.741 & +15:47:01.223 & $95.6^{+7.2}_{-5.8}$  & $0.099^{+0.029}_{-0.033}$ & $1.88^{+0.18}_{-0.21}$ & $0.129^{+0.007}_{-0.008}$ & $6.16^{+0.38}_{-0.38}$  & $17.65^{+0.25}_{-0.25}$ \\ \addlinespace
    \textit{F770W}  & 01:36:41.733 & +15:47:01.294 & $86.6^{+4.3}_{-4.5}$  & $0.296^{+0.034}_{-0.033}$ & $0.63^{+0.08}_{-0.08}$ & $0.173^{+0.007}_{-0.007}$ & $8.25^{+0.36}_{-0.36}$  & $18.53^{+0.02}_{-0.02}$ \\
    \textit{F1000W} & 01:36:41.733 & +15:47:01.245 & $93.2^{+5.0}_{-5.6}$  & $0.395^{+0.083}_{-0.045}$ & $1.45^{+0.71}_{-0.64}$ & $0.250^{+0.015}_{-0.022}$ & $11.9^{+1.0}_{-1.0}$    & $18.78^{+0.15}_{-0.15}$ \\
    \textit{F1300W} & 01:36:41.733 & +15:47:01.247 & $93.1^{+5.0}_{-3.9}$  & $0.393^{+0.071}_{-0.045}$ & $1,46^{+0.63}_{-0.40}$ & $0.251^{+0.009}_{-0.015}$ & $11.99^{+0.72}_{-0.72}$ & $18.82^{+0.08}_{-0.08}$ \\
    \textit{F2100W}\tablenotemark{{\small c}} & 01:36:41.728 & +15:47:01.231 & -{-}                  & -{-}                      & -{-}                   & -{-}                      & -{-}                    & $19.51^{+0.68}_{-0.68}$ \\
  \enddata
  \tablenotetext{{\small \textit{a}}}{Apparent magnitude in the AB-magnitude system. The values are corrected for extinction.}
  \tablenotetext{{\small \textit{b}}}{Uncertainties were determined based on the assumption that the parameter distribution is Gaussian.}
  \tablenotetext{{\small \textit{c}}}{No fit to the data succeeded if PSF convolution was enabled. To determine the central position of the NSC and the magnitude, the data were fit without PSF convolution.}
\end{deluxetable*}

\bibliography{bibliography,phangs_jwst}
\bibliographystyle{aasjournal}



\end{document}